\documentclass[12pt]{article}
\usepackage{hyperref}
\hypersetup{
colorlinks=true,
linkcolor=black,
citecolor=black,
urlcolor=cyan,
filecolor=black,
pdfborder={0 0 0},
}
\def\Red{} 
\def\Gold{\Goldenrod}
\def\Black{} 
\def\Green{} 
\def\RedViolet{} 
\def\Salmon{} 
\def\Goldenrod{} 
\def\RoyalBlue{} 
\usepackage{hyperref}
\hypersetup{
colorlinks=true,
linkcolor=black,
citecolor=black,
urlcolor=cyan,
filecolor=black,
pdfborder={0 0 0},
}
\usepackage{amssymb}
\usepackage{amsfonts}
\usepackage{mathrsfs}
\usepackage{latexsym}
\usepackage{amsmath}
\usepackage{graphics}
\usepackage{graphicx}
\usepackage{sectsty}
\usepackage{setspace}
\usepackage{multirow}
\sectionfont{\normalsize\bf}
\subsectionfont{\rm\normalsize}
\def\be{\begin{equation}}
\def\ee{\end{equation}}
\input amssym.def
\input amssym
\def\sss{\scriptscriptstyle}

\def\Oc{{\cal O}}

\def\C{{\cal C}}
\def\M{{\cal M}}

\def\bpi{{\overline\pi}}

\newcount\nx
\newcount\nxo
\newcount\ny
\newcount\nyo
\newcount\nxd
\newcount\nxod
\newcount\nyd
\newcount\nyod
\newcount\nl

\def\grid#1#2#3#4{
\nxo=#1
\nyo=#2
\nx=#3
\ny=#4    
\multiply \nxo by 4
\multiply \nyo by 4
\advance \nx by 1
\multiply \ny by 4
\linethickness{0.075mm}
\multiput(\nxo,\nyo)(4,0){\nx}{\line(0,1){\ny}}
\nx=#3
\ny=#4
\advance \ny by 1
\multiply \nx by  4
\multiput(\nxo,\nyo)(0,4){\ny}{\line(1,0){\nx}}
}

\def\dgrid#1#2#3#4{
\nxo=#1
\nyo=#2
\nx=#3
\ny=#4
\multiply \nxo by 4
\multiply \nyo by 4
\advance \nxo by - 2
\advance \nyo by - 2
\advance \nx by 1
\multiply \ny by  4
\linethickness{0.3mm}
\multiput(\nxo,\nyo)(4,0){\nx}{\line(0,1){\ny}}
\nx=#3
\ny=#4
\advance \ny by 1
\multiply \nx by  4
\multiput(\nxo,\nyo)(0,4){\ny}{\line(1,0){\nx}}
}

\def\puti#1#2#3{
\nxo=#1
\nyo=#2
\nl=4
\multiply \nxo by \nl
\multiply \nyo by \nl
\divide \nl by 2
\advance \nxo by - \nl
\advance \nxo by - 1
\advance \nyo by - \nl
\advance \nyo by - 1
\put(\nxo,\nyo){$#3$}
}

\def\putl#1#2#3{
\nxo=#1
\nyo=#2
\nl=4
\multiply \nxo by \nl
\multiply \nyo by \nl
\divide \nl by 2
\advance \nxo by - \nl
\advance \nyo by - \nl
\put(\nxo,\nyo){$#3$}
}
\def\puto#1#2{
\nxo=#1
\nyo=#2
\nl=4
\thicklines
\multiply \nxo by \nl
\multiply \nyo by \nl
\divide \nl by 2
\advance \nxo by - \nl
\advance \nyo by - \nl
\put(\nxo,\nyo){\circle{1.5}}
}
\def\case#1#2#3#4{
\nxo=#1
\nyo=#2
\nx=#3
\ny=#4
\nl=4
\multiply \nxo by \nl
\multiply \nyo by \nl
\advance \nxo by - \nl
\advance \nyo by - \nl
\multiply \ny by  \nl
\linethickness{0.3mm}
\multiply \nx by \nl
\multiput(\nxo,\nyo)(\nx,0){2}%
{\line(0,1){\ny}}
\nx=#3
\ny=#4
\multiply \nx by  \nl
\multiply \ny by  \nl
\multiput(\nxo,\nyo)(0,\ny){2}%
{\line(1,0){\nx}}
}
\def\putc#1#2{
\nxo=#1
\nyo=#2
\multiply \nxo by 4
\multiply \nyo by 4
\divide \nl by 2
\advance \nxo by - 2
\advance \nyo by -2
\thicklines
\put(\nxo,\nyo){\circle{1.5}}
\put(\nxo,\nyo){\circle{0.8}}
}
\def\putcc#1#2{
\nxo=#1
\nyo=#2
\nl=4
\multiply \nxo by \nl
\multiply \nyo by \nl
\divide \nl by 2
\advance \nxo by - \nl
\advance \nyo by - \nl
\thicklines
\put(\nxo,\nyo){\circle{2.0}}
\put(\nxo,\nyo){\circle{1.5}}
\put(\nxo,\nyo){\circle{0.8}}
}
\def\const#1#2{
\nxo=#1
\nyo=#2
\multiply \nxo by 4
\multiply \nyo by 4
\advance \nxo by - 2
\advance \nyo by - 2
\put(\nxo,\nyo){\circle*{2.7}}
}

\def\constx#1#2{
\setlength{\unitlength}{0.3mm}
\nxo=#1
\nyo=#2
\linethickness{0.8mm}
\multiply \nxo by 20
\multiply \nyo by 20
\advance \nxo by - 19%
\advance \nyo by - 18%
\put(\nxo,\nyo){\line(1,0){18}}%
\advance \nyo by 16
\put(\nxo,\nyo){\line(1,0){18}}
\advance \nyo by -17
\advance \nxo by 1
\put(\nxo,\nyo){\line(0,1){18}}
\advance \nxo by 16
\put(\nxo,\nyo){\line(0,1){18}}
\setlength{\unitlength}{1.5mm}
}

\numberwithin{equation}{section}

\begin{document}

\thispagestyle{empty}
\begin{flushright}
\end{flushright}
\baselineskip=16pt
\vspace{.5in}
{
\begin{center}
{\bf Complete Equivalence Between Gluon Tree Amplitudes}

{\bf in Twistor String Theory 
and  in Gauge Theory}
\end{center}}
\vskip 1.1cm
\begin{center}
{Louise Dolan}
\vskip5pt

\centerline{\em Department of Physics}
\centerline{\em University of North Carolina, Chapel Hill, NC 27599} 
\bigskip
\bigskip        
{Peter Goddard}
\vskip5pt

\centerline{\em Institute for Advanced Study}
\centerline{\em Princeton, NJ 08540, USA}
\bigskip
\bigskip
\bigskip
\bigskip
\end{center}

\abstract{\noindent 
The gluon tree amplitudes of open twistor string theory, defined as contour integrals 
over the ACCK link variables, are shown to satisfy the BCFW relations, thus confirming that they coincide with the 
corresponding amplitudes in gauge field theory. In this approach, the integration contours are specified
as encircling the zeros of certain constraint functions that force the appropriate relation between the link
variables and the twistor string world-sheet variables. To do this, methods for calculating the tree amplitudes 
using link variables are developed further including diagrammatic methods for organizing and performing the
calculations.

\bigskip

\setlength{\parindent}{0pt}
\setlength{\parskip}{6pt}

\setstretch{1.05}
\vfill\eject
\vskip50pt
\section{\bf Introduction}

In this paper we demonstrate that all gluon tree amplitudes in open twistor string theory \cite{W}--\cite{BW} are identical with the corresponding amplitudes in $N=4$ super Yang-Mills theory. To do this we develop further the approach we used in \cite{DG3,DGsh} to evaluate amplitudes in twistor string theory using the link variables of Arkani-Hamed, Cachazo, Cheung and Kaplan (ACCK) \cite{ACCK1, ACCK2} (see also \cite{SPL, Nandan:2009cc, Bourjaily:2010kw}). In particular, in \cite{DGsh}, we explicitly evaluated all split-helicity tree amplitudes in twistor string theory and showed that this class of amplitudes coincides with the corresponding amplitudes in Yang-Mills theory, by using the Britto-Cachazo-Feng-Witten (BCFW) relations \cite{BCF,BCFW}, which effectively determine all gluon tree amplitudes in Yang-Mills theory in terms of three point functions.

The proof of the BCFW relations \cite{BCFW} proceeds by  making a shift, linear in a complex variable $z$, in two of the momenta on which the amplitude depends and writing the amplitude as a sum over residues of its poles as a function of $z$. We use the fact that this shift is equivalent to a shift by $z$ in just one of the link variables, to analyze poles of the contour integral expression for the twistor string tree amplitude, viewed as a function of $z$, identifying the location of the poles and calculating the residues to produce exactly the terms in the BCFW relation. 

We begin section \ref{amps} by reviewing the introduction of the link variables $c_{ir}$ and the how the general twistor string gluon tree amplitude can be expressed as a contour integral over these variables with the contour drawn precisely to encircle the poles provided by zeros of certain constraint functions, sextic  in the $c_{ir}$, in the denominator of the integrand. (The constraint functions reflect the relation of the link variables to the independent world-sheet variables describing the twistor string.) The contour must include the complex zeros of the constraints in order to produce the tree amplitudes of gauge theory, which are rational functions of the momenta (restricting to just the real roots would yield an irrational amplitude in general), but exclude roots where any of the $c_{ir}$ are zero, which are `spurious' contributions \cite{DG3, SPL, RMV3}.

A proof of the BCFW relation for tree amplitudes in twistor string theory at the level of the string path integral has been given by Skinner  \cite{Skinner:2010cz}, following \cite{Mason:2009sa}. Here we establish this relation for the twistor string gluon tree amplitudes defined as integrals over link variables over the contours just described. 

In section \ref{FormF}, we discuss the form of the integrand function, $F$, for the general gluon tree amplitude, which we find convenient to write as a function $G$ multiplied by the numerator function, $F^{\hbox{\tiny split}}$, for the corresponding split-helicity amplitude. We give simple rules for writing $G$ down in the general case in terms of a line that snakes through a tabular graph whose rows are labeled by the positive helicities and whose columns are labeled by the negative felicities. The poles in $F$ can  be characterized in terms of the `snake' line. 

Using these results, we list in section \ref{PolesM} the points in the space of link variables that correspond to poles of the amplitude, dividing these into four classes labeled (A), (B), (C) and (D), and determining the corresponding location of the pole in $z$. This enables us to state in our notation the BCFW relation that we are trying to prove. In sections \ref{TypeAB}, \ref{TypeC} and \ref{TypeD}, we calculate the contributions of classes (A) and (B), of class (C), and of class (D), respectively, showing that these provide exactly all the terms in the BCFW relation (\ref{BCFWE}). Then, in section \ref{Abs}, we complete the proof of the relation by confirming that there are no other residues contributing to it. The results of the paper and their implications are reviewed in section \ref{Conc}. The representations of amplitudes developed here may lead to simple explicit representations and diagrammatic rules for gluon tree amplitudes and approaches to studying loop amplitudes in twistor string theory \cite{DG1}--\cite{Alday}.

In Appendix \ref{RelG} a number of forms for the function $G$ are derived in terms of rules applied to the `snake line', while Appendix  \ref{Jacobians} gives Jacobians necessary to evaluate the various classes of residues; and in Appendix \ref{Diagrams}, we outline diagrammatic methods that help organize and facilitate calculations of twistor string amplitudes.

\section{\bf Twistor String Amplitudes}
\label{amps}

We consider a twistor string tree amplitude for $N$ gluons, with momenta $p_\alpha$ and helicities $\epsilon_\alpha, 1\leq \alpha\leq N$.  We suppose that there are $m$ positive helicity gluons, labeled $i_1,\ldots,i_m$,  and $n$ negative helicity gluons, labeled $r_1,\ldots, r_n$, where the helicities of the same sign are not in general adjacent, and $N=m+n$. Write $I=\{i_1,\ldots,i_m\}$ and $R=\{r_1,\ldots, r_n\}$; and the gluon momenta ${p^a_\alpha}_{\dot a}
=\pi_\alpha^a\bpi_{\alpha\dot a}.$ The link variables $c_{ir},i \in I, r\in R$ satisfy the $2(m+n)$ linear equations 
\be
\pi_i =\sum_{r\in R} c_{ir}\pi_r\qquad
\bpi_r =-\sum_{i \in I}\bpi_i c_{ir}.\label{L1}
\ee
where we have suppressed the spinor indices.
(See \cite{DG1, DG3} for our conventions.)
These equations are not independent because
they imply momentum conservation \cite{ACCK1}, and
for momenta satisfying this consistency condition they provide $2(m+n)-4$ 
constraints on the $mn$ variables $c_{ir}$, leaving 
$N_R=(m-2)(n-2)$ independent degrees of freedom, which can be taken to be 
$c_{i_ar_b}$, $2\leq a\leq m-1$, $2\leq b\leq n-1$. 

In  \cite{DG3}, we showed how to write the general twistor string gluon tree amplitude ,
\begin{align}
M_{mn}^{A_1A_2\ldots A_N}(p_1,p_2,\ldots p_N) &= f^{A_1A_2\ldots A_N}
\; \delta^4\left(\sum_\alpha \; \pi_\alpha\bpi_\alpha\right)
\,\M_{mn}(p_1,p_2,\ldots p_N),
\end{align}
where the sum $\alpha$ is over both positive and negative helicities,
in terms of a contour integral 
\be \M_{mn}=K_{mn}\oint_\Oc F_{mn}(c) \prod_{a=2}^{m-1} \prod_{b=2}^{n-1} 
{dc_{i_ar_b}\over \C_{ab}}.\label{ointhF}\ee
taken around a contour encircling the simultaneous zeros of the $N_R$ constraint functions $\C_{ab}$, which are defined by
\be \C_{ab}\equiv\C^{i_{a-1}i_ai_{a+1}}_{r_{b-1}r_br_{b+1}} = 0,
\qquad 2\leq a\leq m-1,  \quad 2\leq b\leq n-1, \label{defC}\ee
where
\begin{align}
\C^{ijk}_{rst}&=\left|\begin{matrix} c_{is}c_{it}&c_{it}c_{ir}
& c_{ir}c_{is}\cr c_{js}c_{jt}&c_{jt}c_{jr}& c_{jr}c_{js}\cr
c_{ks}c_{kt}&c_{kt}c_{kr}& c_{kr}c_{ks}\cr\end{matrix}\right|\label{C'}\\
&=c_{it}c_{kr}c^{ij}_{rs}c^{jk}_{st}-c_{ir}c_{kt}c^{jk}_{rs}c^{ij}_{st},\label{C''}
\end{align}
where 
\be
c^{ij}_{rs}=c_{ir}c_{js}-c_{is}c_{jr}.
\ee

The $N_R$ constraints (\ref{defC}) imply that the general constraint (\ref{C'}) vanishes for any $i,j,k\in I$, and $r,s,t\in R$. The contour $\Oc$ should exclude any zero of the constraints at which any of the $c_{ir}$ is zero \cite{DG3}.

The functions $F_{mn}(c)$ are simple rational functions of the $c_{ir}$ 
described in section \ref{FormF}, and the constant 
\begin{align}
K_{mn} = \langle r_1,r_n\rangle^{2-m}[i_1,i_m]^{2-n},
\end{align}
where $\langle \alpha, \beta\rangle \equiv \langle \pi_\alpha, \pi_\beta\rangle,\;
[ \alpha, \beta] \equiv [ \bpi_\alpha, \bpi_\beta]$. In (\ref {ointhF}) the variables $c_{i_ar_b}$ for 
$a=1,m$ or $b=1,n$ are determined in terms of terms of the integration variables 
$c_{i_ar_b},\; 2\leq a \leq m-1,\; 2\leq b\leq n-1$, using the momentum constraints (\ref {L1}).
[The constraints of the form (\ref{defC}) were termed `contiguous constraints' in \cite{DGsh} and what we have called $F(c)$ here was termed $\hat F(c)$ there.]

Following \cite{BCFW}, we consider making a shift in momentum by sending 
\be\pi_{i_1}\mapsto \pi_{i_1}(z)=\pi_{i_1}+z\pi_{r_n}, 
\qquad\bpi_{r_n}\mapsto \bpi_{r_n}(z)=\bpi_{r_n}-z\bpi_{i_1},\label{pishift}\ee
and leaving the other $\pi_\alpha, \alpha\ne i_1$ and $\bpi_\beta, \beta\ne r_n$, unchanged,
\be \pi_{\alpha}\mapsto \pi_{\alpha}(z)=\pi_{\alpha}+z\pi_{r_n}\delta_{\alpha i_1}, 
\qquad\bpi_{\beta}\mapsto \bpi_{\beta}(z)=\bpi_{\beta}-z\bpi_{i_1}\delta_{\beta r_n}.\label{pishift2}\ee
 The corresponding amplitude, which we shall denote $\M_{mn}(z)$ is given by (\ref{ointhF}) but where now the variables $c_{i_ar_b}$ are subjected to constraints given by the shifted momenta. If we use $\tilde c_{i_ar_b}$ for the expression for the link variables in the expression for $\M_{mn}(z)$, we have
 \be \M_{mn}(z)=K_{mn}\oint_\Oc F_{mn}(\tilde c) \prod_{a=2}^{m-1} \prod_{b=2}^{n-1} 
{d\tilde c_{i_ar_b}\over \C_{ab}(\tilde c)}.\label{ointhFz}\ee
 where
 \be
\pi_i +z\pi_{r_n}\delta_{ii_1} =\sum_{r\in R}\tilde c_{ir}\pi_r\qquad
\bpi_r -z\bpi_{i_1}\delta_{rr_n} =-\sum_{i \in I}\bpi_i \tilde c_{ir}.\label{shiftL1}
\ee
We can shift the dependence on $z$ from the momentum constraints into the integrand by noting that if we write $c_{ir} = \tilde c_{ir} -z \delta_{ii_1}\delta_{rr_n} $ the shifted constraints (\ref{shiftL1}) become just the original momentum constraints (\ref{L1}). Thus, writing 
\be 
c_{ir}(z)\equiv \tilde c_{ir}= c_{ir} +z \delta_{ii_1}\delta_{rr_n},\label{shiftc}
\ee
 \be \M_{mn}(z)=K_{mn}\oint_\Oc F_{mn}(c(z)) \prod_{a=2}^{m-1} \prod_{b=2}^{n-1} 
{d c_{i_ar_b}\over \C_{ab}( c(z))},\label{ointhFz2}\ee
where the variables $c_{ir}$ (rather than $c_{ir}(z)$) are subject to the momentum constraints (\ref{L1}). 
In other words, $\M_{mn}(z)$ can be computed from (\ref{ointhF}) by just making the simple shift 
\be 
 c_{i_1r_n}\mapsto c_{i_1r_n} +z \label{shiftc2}
\ee
throughout the integrand (but not the momentum constraints) and leaving all the other $c_{ir}$ unchanged.

The rational function $F(c)$, defined in section \ref{FormF}, does not involve $c_{i_1r_n}$ in its numerator (and only rarely in its denominator) and the only constraint function to involve $c_{i_1r_n}$  is $C_{2,n-1}$, which is linear in the variable $z$. In consequence, $\M_{mn}(z)$ decreases as $z\rightarrow\infty$ at least as fast as $z^{-1}$, and, since it will be seen to be meromorphic, we can write it as the sum of the residues $ \M_{mn}^{z_i}$ at its poles $z_i$,
\be
\M_{mn}(z) 
= \sum_{z_i} {\M^{z_i}_{mn}\over z - z_i}\qquad
\hbox{so that}\qquad
\M_{mn} =\M_{mn}(0) 
= -\sum_{z_i} {1\over z_i}\M^{z_i}_{mn}.
\label{exppole}\ee
We shall show that, as would be hoped, the terms in this expression for $\M_{mn}$ are just those in the BCFW relation for gluon tree amplitudes in Yang-Mills theory, so that, given the equalities already established for three-point functions, the gluon tree amplitudes are the same in twistor string theory and Yang-Mills theory.

\section{\bf  The Form of $F_{mn}(c)$}
\label{FormF}
The form of $F_{mn}(c)$ for a general gluon tree amplitude was calculated in \cite{DG3} but using a different form of the constraint functions $\C_{ab}$ from that specified by (\ref{defC}). The form appropriate to the choice (\ref{defC}) can be calculated using the method employed in  \cite{DGsh} to calculate it for contiguous constraints in the split-helicity case, where all the helicities of the same sign are adjacent. Alternatively, the calculation may be performed more quickly, by noting that the ratio of the functions $F_{mn}(c)$ appropriate for two different choices of constraints is independent of the order of the helicities. Thus, for a given order of helicities, we can obtain $F_{mn}(c)$ by multiplying the form given in \cite{DG3} by the ratio of the  function $F^{\hbox{\tiny split}}_{mn}(c)$ for split-helicity amplitudes appropriate to contiguous constraints to the corresponding form for split-helicity amplitudes given in \cite{DG3}. Thus 
\be F_{mn}(c) = G_{mn}(c) F^{\hbox{\tiny split}}_{mn}(c)\label{GF}\ee
where $G_{mn}(c)$ is the function for the amplitude divided by that for the split-helicity amplitude, which can be calculated using any choice of constraints. 

From equation (13) of \cite{DGsh}, we have 
\be F^{\hbox{\tiny split}}_{mn}(c)={1\over c^{i_1i_{2}}_{r_1r_2}c^{i_{m-1}i_m}_{r_{n-1}r_n}}
\prod_{a=2}^{m-2}\prod_{b=2}^{n-2}c^{i_{a}i_{a+1}}_{r_{b}r_{b+1}}c_{i_ar_{b}}
c_{i_{a+1}r_{b+1}}\prod_{a=2}^{m-2} 
c_{i_ar_1}c_{i_{a+1}r_n}\prod_{b=2}^{n-2} c_{i_1r_{b}}c_{i_mr_{b+1}}
\prod_{a=2}^{m-1}\prod_{b=2}^{n-1} c_{i_{a}r_{b}}.\label{Fsplit}\ee

Consider an amplitude, 
with $m$ positive helicities and $n$ negative helicities, where $i_1=1$ and $r_n=m+n$, with the first $m_1$ helicities being positive, the next $n_1$ being negative, the next $m_2-m_1$ being positive, the next $n_2-n_1$ being negative, and so on until we end with $m_p-m_{p-1}$ positive helicities followed by $n_p-n_{p-1}$ negative heliciites. Thus $p$ is the number of strings of adjacent positive helicities, and so also the number of strings of adjacent negative helicities, $m_p=m, n_p=n$ and we write $m_0=n_0=0$. Then, from \cite{DG3}, 
\be
G_{mn}(c)={1\over c_{i_mr_1}}\prod_{e=1}^pc_{i_{m_e}r_{n_{e-1}+1}}
\prod_{e=1}^{p-1}{c^{i_{m_e}i_{m_e+1}}_{r_{n_e}r_{n_e+1}}
\over c_{i_{m_e}r_{n_e}}c_{i_{m_e}r_{n_e+1}}c_{i_{m_e+1}r_{n_e+1}}}.\label{defG}
\ee
We have assumed that the first helicity is positive and the last helicity is negative. We can always arrange that this is the case by taking a suitable starting point for numbering the gluons. We give a proof of (\ref{defG}) in Appendix \ref{RelG} and also expressions for $G_{mn}(c)$ that begin or end with other helicities. 

We can label this amplitude by a diagram consisting of a rectangular array of squares with $m$ rows and $n$ columns, labeled by the positive and negative helicities, on which we draw a line which snakes from the top left hand corner to the bottom right hand corner, first taking $m_1$ steps
downwards, then $n_1$ steps to the right, followed by $m_2-m_1$ steps downward and then $n_2-n_1$ to the right and so on until the bottom right hand corner is reached with $n-n_{p-1}$ steps to the right. Such a diagram has corners at points with coordinates $(m_e, n_{e-1}),\;1\leq e\leq p$, relative to axes pointing downwards and to the right, and at points with coordinates $(m_e, n_e),\;1\leq e\leq p-1$. With a corner of the first type we associate a factor of $c_{i_{m_e}r_{n_{e-1}+1}}$, 
and with a corner of the second type we associate a factor of $c^{i_{m_e}i_{m_e+1}}_{r_{n_e}r_{n_e+1}}/ c_{i_{m_e}r_{n_e}}c_{i_{m_e}r_{n_e+1}}c_{i_{m_e+1}r_{n_e+1}}$. $G_{mn}(c)$ is the product of these factors divided by $c_{i_mr_1}$.
\setlength{\unitlength}{1.5mm}
\vskip45truemm
\hbox to \hsize{\hfil
\hbox{\hskip-25truemm
\begin{picture}(1,1)
\grid{-1}{-1}{8}{7}
\put(-7,21){$i_1$}\put(-7,17){$i_2$}\put(-7,13){$i_3$}\put(-7,9){$i_4$}
\put(-7,5){$i_5$}\put(-7,1){$i_6$}\put(-7,-3){$i_7$}
\put(-3,25){$r_1$}\put(1,25){$r_2$}\put(5,25){$r_3$}\put(9,25){$r_4$}
\put(13,25){$r_5$}\put(17,25){$r_6$}\put(21,25){$r_7$}\put(25,25){$r_8$}
\RedViolet
\linethickness{0.5mm}
\put(-4,12){\line(1,0){16}}\put(-4,12){\line(0,1){12}}
\put(12,4){\line(1,0){12}}\put(12,4){\line(0,1){8}}
\put(24,-4){\line(0,1){8}}\put(24,-4){\line(1,0){4}}
\Salmon
\put(-4,16){\circle*{1.5}}
\put(0,12){\circle*{1.5}}\put(4,12){\circle*{1.5}}\put(8,12){\circle*{1.5}}\put(12,12){\circle*{1.5}}
\put(12,8){\circle*{1.5}}
\put(12,4){\circle*{1.5}}\put(16,4){\circle*{1.5}}\put(20,4){\circle*{1.5}}
\Green
\put(4,12){\circle{2.0}}\put(8,12){\circle{2.0}}\put(12,12){\circle{2.0}}
\put(12,8){\circle{2.0}}
\put(12,4){\circle{2.0}}\put(16,4){\circle{2.0}}\put(20,4){\circle{2.0}}\put(24,4){\circle{2.0}}
\put(24,0){\circle{2.0}}
\Black
\put(6,-8){Figure 1}
\end{picture}}\hfil}
\vskip15truemm
Figure 1 illustrates this diagram for a 15-point amplitude with helicities 
\be
(+,+,+,-,-,-,-,+,+,-,-,-,+,+,-)\label{15pt}
\ee
for which $p=3$. 

The contributions to the BCFW recursion for the Yang-Mills tree amplitude with these helicities can be labeled by points on the `snake' line. The contribution associated with a point $\alpha$ steps along the `snake' corresponds to the product of gluon tree amplitudes with helicities $(\epsilon_1,\ldots,\epsilon_\alpha,\epsilon)$ and 
$(-\epsilon,\epsilon_{\alpha+1},\ldots,\epsilon_N)$, for $\epsilon=\pm1$; we shall refer to those 
 with $\epsilon=-1$ as negative helicity contributions and those with $\epsilon=+1$ as positive helicity contributions. Not all of these contributions are nonzero. In the case illustrated in Figure 1, the nonzero negative helicity contributions have been indicated by  solid (salmon) circles  and the nonzero positive helicity contributions by open (green) circles. More generally, the point $(m',n')$ on the `snake' corresponds to a nonzero negative helicity contribution if 
\be 
(m',n')= (2,0)\;\hbox{or}\; (m,n-2)\quad \hbox{or}\quad 
2\leq m'\leq m-1\;\hbox{and} \;1\leq n'\leq n-2 \label{negh}
\ee
and to a nonzero positive helicity contribution if 
\be 
(m',n')= (1,1)\;\hbox{or}\; (m-1,n-1)\quad \hbox{or}\quad 
1\leq m'\leq m-2\;\hbox{and} \;2\leq n'\leq n-1.  \label{posh}
\ee
We shall show that we get precisely these contributions for the twistor string tree amplitudes in the expansion (\ref{exppole}).

In general, the factors in the denominator of $G_{mn}(c)$ (\ref{defG}), other than $c_{i_mr_1}$,  are canceled by factors in the numerator of $F^{\hbox{\tiny split}}_{mn}(c)$ (\ref{Fsplit}) so, typically, the denominator of $F_{mn}(c)$ is $c_{i_mr_1}c^{i_1i_2}_{r_1r_2}c^{i_{m-1}i_m}_{r_{n-1}r_n}$. More specifically, the 
singularities of $F_{mn}(c)$ are at:
\begin{alignat}{3}\label{rpoles}
\hbox{(a)}\quad & c^{i_1i_2}_{r_1r_2}=0\quad & \hbox{unless} \quad &m_1=n_1=1; \cr
\hbox{(b)}\quad & c^{i_{m-1}i_m}_{r_{n-1}r_n}=0\quad & \hbox{unless} \quad &m_{p-1}=m-1, n_{p-1}=n-1;\cr
\hbox{(c)}\quad & c_{i_mr_1}=0\quad & \hbox{unless} \quad &p=1;
\end{alignat}
\vskip-6pt
and, exceptionally, they are at:
\begin{alignat}{3}\label{spoles}
\hbox{(d)}\quad & c_{i_1r_{n-1}}=0\quad & \hbox{if} \quad &m_1=1\quad \hbox{and}\quad n_1= n-2\quad\hbox{or}\quad n-1; \cr
\hbox{(e)}\quad & c_{i_2r_{n}}=0\quad & \hbox{if} \quad &m_{p-1}=1\quad \hbox{or}\quad 2\quad \hbox{and}\quad n_{p-1}=n-1;\cr
\hbox{(f)}\quad & c_{i_1r_{n}}=0\quad & \hbox{if} \quad &p=2,\quad  m_1=1,\quad  n_1=n-1,\quad  m_2=m\quad  \hbox{and}\quad  n_2=n.
\end{alignat}

\section{\bf  The Poles in $\M_{mn}(z)$}
\label{PolesM}

For any value of $z$, $\M_{m,n}(z)$, as defined by (\ref{ointhFz2}), 
\be \M_{mn}(z)=K_{mn}\oint_\Oc F_{mn}(c(z)) \prod_{a=2}^{m-1} \prod_{b=2}^{n-1} 
{d c_{i_ar_b}\over \C_{ab}( c(z))},\nonumber\ee
receives contributions from the points which are the common solutions of the $N_R=(m-2)(n-2) $ constraints $\C_{ab}(c(z))=0$ and the momentum constraints (\ref{L1}), which provide the $2(m+n)-4$ additional conditions necessary to determine all the $mn$ variables $c_{i_ar_b}$.
The contour $\Oc$ is defined as encircling these solutions but excluding ones at which, for generic momenta, any of the $c_{i_ar_b}$ vanish, which correspond to spurious solutions \cite{DG3}.

Poles of $\M_{m,n}(z)$ in $z$ will arise when, as $z$ varies, the contour $\Oc$ is pinched between a pole of the integrand, corresponding to one of the factors, $d(z)$, say, in the denominator of $F_{mn}(c(z))$, and the poles corresponding to the vanishing of the constraints $\C_{ab}$. The residue of such a pole, at $z=z_i$, say, will be given by 
 \be \hbox{Res}_{z_i}\M_{mn}=K_{mn}\oint_{\overline\Oc} F_{mn}(c(z)) \prod_{a=2}^{m-1} \prod_{b=2}^{n-1} 
{d c_{i_ar_b}\over \C_{ab}( c(z))}dz,\label{ResM}\ee
where the contour $\overline{\Oc}$ encircles the solutions of 
\be
d(z) =0;\qquad \C_{ab}(z)=0, \quad 2\leq a\leq m-1, \;2\leq b\leq n-1,
\label{csol}\ee
at $z=z_i$. The locations, $z_i$, of the poles are determined by the intersection of the ($2(m+n)-4$)-dimensional surfaces of solutions to (\ref{csol}) with the hyperplane specified by the momentum constraints (\ref{L1}). [Note that, although zeros of the constraints at which one of the $c_{ir}$ vanishes are excluded from $\Oc$ at $z=0$, zeros of the $c_{ir}$ can develop at the pinch corresponding to a pole at $z=z_i$, and we shall see that this is typically the case.] In section \ref{Abs}, we analyze the possible ($2(m+n)-4$)-dimensional surfaces of solutions to (\ref{csol}). Here we discuss the general form of what results. 

(A) If we consider the pole in the integrand given by (\ref{rpoles}(a)), $c^{i_1i_2}_{r_1r_2}=0$, the corresponding solutions satisfy
\be c^{i_1i_2}_{r_br_{b+1}}(z)=0, \quad 1\leq b\leq n-1; \qquad \C_{ab}=0, \quad 3\leq a\leq m-1,\;
2\leq b\leq n-1.\label{caseA}\ee

(B) Similarly, if we consider the pole given by (\ref{rpoles}(b)), $ c^{i_{m-1}i_m}_{r_{n-1}r_n}=0$, the corresponding solutions satisfy
\be  c^{i_{a}i_{a+1}}_{r_{n-1}r_n}(z)=0, \quad 1\leq a\leq m-1; \qquad \C_{ab}=0,\quad 2\leq a\leq m-1,\;
2\leq b\leq n-2.\label{caseB}\ee

(C) If we consider the the pole given by (\ref{rpoles}(c)), $c_{i_mr_1}=0$, there is a range of potential solutions, labeled by integers $(m',n')$, $2\leq m'\leq m-1,\; 1\leq n'\leq n-2$, satisfying
\begin{align}
  c_{i_ar_b}=0, \quad m'< a\leq m,\; 1\leq b\leq n' ;\qquad &c^{i_{a}i_{a+1}}_{r_{b}r_{b+1}}(z)=0,
\quad 1\leq a < m',\; n'<b<n;\cr
\C_{ab}=0, \quad 2\leq a\leq m'-1,\;2\leq b\leq n'\quad\hbox{or}&\quad m'< a\leq m-1,\;n'+2\leq b\leq n-1.&\label{caseC}
\end{align}

Not all these solutions, (A), (B), (C), will give nonzero residues for a particular amplitude. We shall see those of classes (A), (B) or (C) will if  the `snake' line associated with the amplitude, defined in section  \ref{FormF}, contains the point $(2,0), (m, n-2)$ or $(m',n')$, respectively. These residues correspond precisely to the negative helicity contributions to the BCFW relations. 

(D) There is a fourth class of potential solutions, 
labeled by integers $(m',n')$, $1\leq m'\leq m-2,\; 2\leq n'\leq n-1$, or $(m',n')=(1,1)$ or $(m-1,n-1)$, satisfying
\begin{align}
  c_{i_ar_b}(z)=0, \quad 1\leq a\leq m',\; n'< b\leq n ;\qquad &c^{i_{a}i_{a+1}}_{r_{b}r_{b+1}}=0,
\quad m'< a < m,\; 1\leq b<n';\cr
\C_{ab}=0, \quad 2\leq a\leq m',\;2\leq b\leq n'-1\quad\hbox{or}&\quad m'< a\leq m-1,\;n'+1\leq b\leq n-1.&\label{caseD}
\end{align}
The solutions in this class will give nonzero residues if the `snake' line contains the point $(m',n')$ and they correspond precisely to the positive helicity contributions to the BCFW relations as listed in section \ref{FormF}. They correspond to residues associated with pinches with the poles given by (\ref{spoles}(d,e,f)) and also spurious contributions. 

Each of these four classes of solutions fit into the following framework. We divide the positive indices $I$ into two complementary subsets $I_1, I_2$, and $R$ into two complementary subsets $R_1, R_2$ and impose the following conditions
\begin{align}
  c_{ir}(z)=0, \quad i\in I_2,\; r\in R_1 ;\qquad &c^{ij}_{rs}(z)=0,
\quad \quad i,j\in I_1,\; r,s\in R_2;\label{genf1}\\
\C^{ijk}_{rst}=0, \quad i,j,k\in I_1,\;r,s,t\in &\overline R_1\quad\hbox{or}\quad i,j,k\in \bar I_2,\;r,s,t\in R_2,&\label{genf2}
\end{align}
where $\overline R_1$ is defined by appending to $R_1$ a point of $R_2$ and $\bar I_2$ is defined by appending to $I_2$ a point of $I_1$. We stipulate that {\it either} $i_1\in I_2$ and $r_n\in R_1$ {\it or} $i_1\in I_1$ and $r_n\in R_2$, so that one of the conditions (\ref{genf1}) involves $z$ (but none of those in (\ref{genf2}) do).  The two sets of conditions $\C^{ijk}_{rst}=0$ in (\ref{genf2}) contain $(m'-2)(n'-1)$ and $(m''-1)(n''-2)$ conditions respectively, where $m', n',m'', n''$ 
denote the number of elements of $I_1, I_2, R_1, R_2,$ respectively, so that $m'+m''=m$ and $n'+n''=n$. The total number of independent conditions provided by (\ref{genf1}) and (\ref{genf2}) is $N_R+1$.

The equations (\ref{genf1}) determine the value of $z$ at which any pole associated with these conditions occurs. It follows from equations (\ref{pishift2}), (\ref{shiftL1}) and (\ref{shiftc}) that
\be
\pi_i(z) =\sum_{r\in R} c_{ir}(z)\pi_r(z),\qquad
\bpi_r(z) =-\sum_{i \in I}\bpi_i(z) c_{ir}(z).\label{L1z}
\ee
Let us consider the implications of the conditions (\ref{genf1}) for these equations, for convenience temporarily suppressing  the dependence on $z$. The conditions $c^{ij}_{rs}=0$,
$ i,j\in I_1,\; r,s\in R_2$ enable us to write
\be
c_{ir}=\lambda_i\mu_r,\quad i\in I_1,\;r\in R_2,\qquad\hbox{for some}\quad \lambda_i, \mu_r.
\label{LM}\ee
The relations (\ref{L1z}) become:
\begin{alignat}{7}
\hbox{for}\;&i\in I_1,\quad&\pi_i &=\sum_{r\in R_1} c_{ir}\pi_r+c_{ir_0}\pi_{r_0},
&&\hbox{where}\; \; c_{ir_0}=\lambda_i,\quad\pi_{r_0}=\sum_{r\in R_2} \mu_{r}\pi_r;\cr
\hbox{for}\;&r\in R_1,\quad&\bpi_r &=-\sum_{i \in I_1}\bpi_ic_{ir},
&&\hskip-10pt\hbox{and set}\; \; \bpi_{r_0} =-\sum_{i \in I_1}\bpi_i\lambda_i=-\sum_{i \in I_1}\bpi_ic_{ir_0};\cr
\hbox{for}\;&r\in R_2,\quad&\bpi_r &=-\bpi_{i_0}c_{i_0r}-\sum_{i \in I_2}\bpi_ic_{ir},\; \;
&&\hbox{where}\; \; c_{i_0r}=\mu_r,\quad\bpi_{i_0}=\sum_{i \in I_1}\bpi_i\lambda_{i};\cr
\hbox{for}\;&i\in I_2,&\pi_i &=\sum_{r\in R_2} c_{ir}\pi_r,
&&\hskip-10pt\hbox{and set}\; \; \pi_{i_0} =\sum_{r\in R_2} \mu_r\pi_r=\sum_{r\in R_2} c_{i_0r}\pi_r.
\label{IR}
\end{alignat}
So, if we adjust the definitions above by writing $\overline R_1= R_1\cup\{r_0\}$ and $\bar I_2=I_2\cup\{i_0\}$, the conditions (\ref{L1z}) are satisfied with $I$ and $R$ replaced with $I_1$ and $\bar R_1$, respectively, or with $\bar I_2$ and $ R_2$, respectively. Note that $\pi_{i_0}=\pi_{r_0}$ and 
$\bpi_{i_0}=-\bpi_{r_0}$. 

In the cases (A), (B) and (C), 
\be 
I_1=\{i_1,\ldots,i_{m'}\},\quad R_1=\{r_1,\ldots,r_{n'}\},\quad
I_2=\{i_{m'+1},\ldots,i_{m}\},\quad R_2=\{r_{n'+1},\ldots,r_{n}\},\label{IRC}
\ee
with $R_1=\O$ in case (A) and $I_2=\O$ in case (B). 
In these cases, the conditions associated with  $I_1,\bar R_1$ are those appropriate to an amplitude with helicities $(m',n'+1)$ and the conditions associated with  $\bar I_2,R_2$ are those appropriate to an amplitude with helicities $(m''+1,n'')$, as is appropriate for negative helicity contributions.

In the case (D), 
\be 
I_2=\{i_1,\ldots,i_{m'}\},\quad R_2=\{r_1,\ldots,r_{n'}\},\quad
I_1=\{i_{m'+1},\ldots,i_{m}\},\quad R_1=\{r_{n'+1},\ldots,r_{n}\}.
\ee
In this case, the conditions associated with  $\bar I_2,R_2$ are those appropriate to an amplitude with helicities $(m'+1,n')$ and the conditions associated with  $I_1,\bar R_1$ are those appropriate to an amplitude with helicities $(m'',n''+1)$, as is appropriate for positive helicity contributions.

Because the conditions (\ref{L1z}) guarantee momentum conservation,
\be
\sum_{i \in I_1}\bpi_i\pi_i+\sum_{r\in R_1} \bpi_r\pi_r=-\bpi_{r_0}\pi_{r_0}=\bpi_{i_0}\pi_{i_0}=
-\sum_{i \in I_2}\bpi_i\pi_i-\sum_{r\in R_2} \bpi_r\pi_r.
\label{momc}
\ee
Restoring the explicit dependence on $z$, we can determine its value  by squaring (\ref{momc}). This gives 
\be 
z=z_A\equiv s_A/[i_1|P_A|r_n\rangle,\label{zPA}
\ee
where 
\be 
s_A=\left(\sum_{\alpha\in A}p_\alpha\right)^2,\qquad 
[i_1|P_A|r_n\rangle=\sum_{\alpha\in A}[i_1,\alpha]\langle\alpha,r_n\rangle
\ee
and $A=I_1\cup R_1$ if $i_1\in I_1$ and $A=I_2\cup R_2$ if $i_1\in I_2$. In each of the four cases (A) to (D) described above, $A=\{i_1,\ldots, i_{m'}, r_1, \dots, r_{n'}\}$, and so, in these cases, as in \cite{BCFW},
\be 
z=z_{m'n'}\equiv s_{m'n'}/[i_1|P_{m'n'}|r_n\rangle\label{zPmn}
\ee
where $s_{m'n'}= (p_{i_1}+\ldots+p_{i_{m'}}+p_{r_1}+\ldots+p_{r_{n'}})^2$ and $P_{m'n'}$ is similarly defined. 

Noting that
\begin{align}
&\pi_{i_a}(z_{m'n'})=\pi_{i_a}+z_{m'n'}\pi_{r_n}\delta_{a1},\quad \bpi_{i_a}(z_{m'n'})=\bpi_{i_a}, \quad 1\leq a\leq m,\cr
&\pi_{r_b}(z_{m'n'})=\pi_{r_b},\quad \bpi_{r_b}(z_{m'n'})=\bpi_{r_b}-z_{m'n'}\bpi_{i_1}\delta_{bn}, \quad 1\leq  b\leq n,\label{ma}
\end{align}
in cases (A), (B) and (C), the momenta associated with the $(m',n'+1)$ amplitude are:
\be
\pi_{i_a}(z_{m'n'}),\bpi_{i_a}(z_{m'n'}), 1\leq a\leq m';\quad
\pi_{r_b}(z_{m'n'}), \bpi_{r_b}(z_{m'n'}),1\leq  b\leq n'; \quad\pi_{r_0},\;\bpi_{r_0};\label{mb}
\ee
 the momenta associated with the $(m''+1,n'')$ amplitude are:
 \be
\pi_{i_0},\;\bpi_{i_0};\quad\pi_{i_a}(z_{m'n'}),\bpi_{i_a}(z_{m'n'}), m'< a\leq m;\quad
\pi_{r_b}(z_{m'n'}), \bpi_{r_b}(z_{m'n'}),n'<  b\leq n; \label{mc}
\ee
in case (D), the momenta associated with the $(m'+1,n')$ amplitude are:
\be
\pi_{i_a}(z_{m'n'}),\bpi_{i_a}(z_{m'n'}), 1\leq a\leq m';\quad
\pi_{r_b}(z_{m'n'}), \bpi_{r_b}(z_{m'n'}),1\leq  b\leq n'; \quad\pi_{i_0},\;\bpi_{i_0};\label{md}
\ee
 the momenta associated with the $(m'',n''+1)$ amplitude are:
 \be
\pi_{r_0},\;\bpi_{r_0};\quad\pi_{i_a}(z_{m'n'}),\bpi_{i_a}(z_{m'n'}), m'< a\leq m;\quad
\pi_{r_b}(z_{m'n'}), \bpi_{r_b}(z_{m'n'}),n'<  b\leq n;\label{me}
\ee
where, by (\ref{momc}), 
\begin{align}
\pi_{i_0}=\pi_{r_0}=\kappa P_{m'n'}|i_1], \quad &\bpi_{i_0}=-\bpi_{r_0}=\mp\bar\kappa P_{m'n'}|r_n\rangle, \label{iPr0}\\
[i_1,i_0]\langle r_0,r_n\rangle = &[i_1|P_{m'n'}|r_n\rangle=(\kappa\bar\kappa)^{-1}. \label{iPr}
\end{align}
where we take the upper sign in (\ref{iPr0}) in cases (A), (B) and (C), and the lower sign in case (D).

The BCFW \cite{BCF,BCFW} relations that we are seeking to prove may now be stated as 
\be 
\M_{mn}=\sum_{(m',n')} {1\over s_{m'n'}}\left[\M_{m',n'+1}\M_{m''+1,n''}+\M_{m'+1,n'}\M_{m'',n''+1}\right]\label{BCFWE}
\ee
where the sum is over the points $(m',n')$ on the `snake' line specified by (\ref{negh}) and (\ref{posh}); and the arguments of $\M_{m',n'+1}$ and $M_{m'+1,n'}$ are $p_{\alpha_i}(z_{m'n'})=\bpi_{\alpha_i}\pi_{\alpha_i}(z_{m'n'}), 1\leq i\leq m'+n'$, together with $\bpi_{r_0}\pi_{r_0}$ in former case and $\bpi_{i_0}\pi_{i_0}$ in the latter case, while the arguments of $\M_{m''+1,n''}$ and $M_{m'',n''+1}$ are $p_{\alpha_i}(z_{m'n'})=\bpi_{\alpha_i}(z_{m'n'})\pi_{\alpha_i}, m'+n'< i\leq N$, together with $\bpi_{i_0}\pi_{i_0}$ in former case and $\bpi_{r_0}\pi_{r_0}$ in the latter case. For a given $(m',n')$, the first term in the sum will be present only if there is a negative helicity contribution corresponding to this point on the `snake' line and the second only if there is a positive helicity contribution. 

\section{\bf Contributions of Types (A) and (B) }
\label{TypeAB}

In the split-helicity case, which we discussed in \cite{DGsh}, there are only two contributions to (\ref{BCFWE}), a negative helicity corresponding to the point $(2,0)$ on the `snake' line, and a positive helicity contribution corresponding to $(m,n-2)$. For a general amplitude, we see from (\ref{negh}) that (\ref{BCFWE}) will contain a contribution corresponding to $(2,0)$  provided that $m_1>1$ and, from (\ref{posh}), a contribution corresponding to $(m,n-2)$ provided that $n_{p-1}<n-1$. In this section, we shall show that, under these conditions,  the type (A) contribution given by (\ref{caseA}) and the type (B) contribution given  by (\ref{caseB}) provide the appropriate terms in (\ref{BCFWE}).

In (\ref{rpoles}(a)), we noted that the denominator of $F_{mn}(c)= G_{mn}(c) F^{\hbox{\tiny split}}_{mn}(c)$ has a factor $c^{i_1i_2}_{r_1r_2}$ unless $m_1=n_1=1$. In the case that it has such a factor, $\M_{mn}(z)$ has a pole at $z=z_{20}$, defined by (\ref{zPmn}), with residue given by (\ref{ResM}),
 \be \hbox{Res}_{z_{20}}\M_{mn}=K_{mn}\oint_{\overline\Oc} F_{mn}(c(z)) \prod_{a=2}^{m-1} \prod_{b=2}^{n-1} 
{d c_{i_ar_b}\over \C_{ab}( c(z))}dz,\nonumber\ee
where the contour $\overline{\Oc}$ encircles  $c^{i_1i_2}_{r_1r_2}=0$ and the zeros of the $\C_{ab}$. From (\ref{C''}),
\be 
\C_{22}=\C^{i_{1}i_2i_{3}}_{r_{1}r_2r_{3}}=c_{i_1r_3}c_{i_3r_1}c^{i_1i_2}_{r_1r_2}c^{i_2i_3}_{r_2r_3}-c_{i_1r_1}c_{i_3r_3}c^{i_2i_3}_{r_1r_2}c^{i_1i_2}_{r_2r_3}=
-c_{i_1r_1}c_{i_3r_3}c^{i_2i_3}_{r_1r_2}c^{i_1i_2}_{r_2r_3}
\ee
at $c^{i_1i_2}_{r_1r_2}=0$, exposing a factor of $c^{i_1i_2}_{r_2r_3}$ in the denominator of $F_{mn}(c(z))$, unless it is cancelled by a similar factor in the numerator. From (\ref{GF}) and (\ref{defG}), there will be such a factor if and only if $m_1=1, n_1=2$. Proceeding iteratively in this way for $2\leq b\leq n-1$, at 
$c^{i_1i_2}_{r_{b-1}r_{b}}=0$,
\be 
\C_{2b}=\C^{i_{1}i_2i_{3}}_{r_{b-1}r_br_{b+1}}=
-c_{i_1r_{b-1}}c_{i_3r_{b+1}}c^{i_2i_{3}}_{r_{b-1}r_b}c^{i_1i_2}_{r_br_{b+1}},\label{C2b}
\ee
producing a factor of $c^{i_1i_2}_{r_{b}r_{b+1}}$ in the denominator of the integrand unless $m_1=1, n_1=b$. If $m_1=1$, then $n_1=b$ for some $b$ with $2\leq b\leq n-1$, so one of the poles at $c^{i_1i_2}_{r_{b-1}r_{b}}=0$ will be cancelled unless $m_1>1$. Note that, for $b=n-1$, the final factor in (\ref{C2b}) should be 
$c^{i_1i_2}_{r_{n-1}r_{n}}(z)$. Then, for $m_1>1$, performing the integrations with respect to $c_{2b}, 2\leq 
b\leq n-1$, and $z$,
 \be \hbox{Res}_{z_{20}}\M_{mn}=-K_{mn}\oint_{\widetilde \Oc}{\widetilde F_{mn}(c)\over c_{i_2r_{n-1} }J^A}  \prod_{a=3}^{m-1} \prod_{b=2}^{n-1} 
{d c_{i_ar_b}\over \C_{ab}( c)},\label{Resz20}\ee
where the contour $\widetilde \Oc$ encircles the zeros of the $\C_{ab}( c)$, $J^{A}$ is the Jacobian of 
$c^{i_1i_2}_{r_1r_2},c^{i_1i_2}_{r_2r_3},\dots, c^{i_1i_2}_{r_{n-2}r_{n-1}}$ with respect to $c_{i_2r_2}, c_{i_2r_3}, \ldots, c_{i_2r_{n-1}}$, when $c^{i_1i_2}_{r_{b}r_{b+1}}=0, 1\leq b\leq n-2$, calculated in Appendix \ref{Jacobians}(b), and
\begin{align}
 \widetilde F_{mn}(c)&=G_{mn}(c) F^{\hbox{\tiny split}}_{mn}(c)c^{i_1i_{2}}_{r_{1}r_2}\prod_{b=2}^{n-1}{1\over c_{i_1r_{b-1}}c_{i_3r_{b+1}}c^{i_2i_{3}}_{r_{b-1}r_b}}\cr
 &=G_{m-1,n}(c) F^{\hbox{\tiny split}}_{m-1,n}(c){1\over c_{i_1r_1}}\prod_{b=1}^{n-1}c_{i_{2}r_{b}}.\label{wtF}
 \end{align}
In (\ref{wtF}), the arguments of $F^{\hbox{\tiny split}}_{m-1,n}(c)$ are $c_{ab}, \;2\leq a\leq m,\;1\leq b\leq n,$
and $G_{m-1,n}(c)=G_{mn}(c)$ does not depend on $c_{1b}, \;1\leq b\leq n,$ as $m_1>1.$

Defining $c^R_{2b}=\mu_{b}, c^R_{ab}=c_{ab}, \;3\leq a\leq m-2,\; 2\leq b\leq n-2,$ where $\mu_b$ is as in (\ref{clm}), then 
$\C_{2b}(c)=\lambda_2^2\C_{2b}(c^R),$ $F^{\hbox{\tiny split}}_{m-1,n}(c)=\lambda_2^{n-4}F^{\hbox{\tiny split}}_{m-1,n}(c^R),$ and $G_{m-1,n}(c)=G_{m-1,n}(c^R).$ Writing $F_{m-1,n}(c^R)=G_{m-1,n}(c^R)F^{\hbox{\tiny split}}_{m-1,n}(c^R)$,

 \be \hbox{Res}_{z_{20}}\M_{mn}={K_{m-1,n}\over \langle r_n, r_0\rangle\lambda_1\lambda_2^2}\oint_{\widetilde \Oc} F_{m-1,n}(c^R)\prod_{a=3}^{m-1} \prod_{b=2}^{n-1} 
{d c_{i_ar_b}^R\over \C_{ab}( c^R)},\ee
where
\be
K_{m-1,n} = \langle r_1,r_n\rangle^{3-m}[i_0,i_m]^{2-n}.
\ee
From (\ref{IR}), $\lambda_1=-[r_0,i_2]/[i_1,i_2],\lambda_2=[r_0,i_1]/[i_1,i_2]$, and using (\ref{iPr}) together with 
$\bpi_{r_0}=-\bpi_{i_0}$,
 \begin{align}
  \hbox{Res}_{z_{20}}\M_{mn}&=-{[i_1,i_2]^3\over [i_1|P_{20}| r_n\rangle[i_1,r_0][r_0,i_2]}K_{m-1,n}\oint_{\widetilde \Oc} F_{m-1,n}(c^R)\prod_{a=3}^{m-1} \prod_{b=2}^{n-1} 
{d c_{i_ar_b}^R\over \C_{ab}( c^R)},\cr
&=-{z_{20}\over s_{20}}\M_{2,1}\M_{m-1,n},
\label{contA}\end{align}
where the arguments of $\M_{2,1}$ and $\M_{m-1,n}$ are the momenta described in (\ref{mb}) and (\ref{mc}), and we obtain the appropriate contribution to BCFW relation (\ref{BCFWE}) whenever $m_1>1$, {\it i.e.} whenever the point $(2,0)$ lies on the `snake' line described in section \ref{FormF}. Similarly, it can be shown that, whenever the point $(m,n-2)$ lies on the `snake' line, 
 \begin{align}
  \hbox{Res}_{z_{m,n-2}}\M_{mn}=-{z_{m,n-2}\over s_{m,n-2}}\M_{m,n-1}\M_{1,2},
\label{contB}\end{align}
provides the appropriate contribution to (\ref{BCFWE}).

\section{\bf Contributions of Type (C) }
\label{TypeC}

We now consider contributions to (\ref{exppole}) arising from poles in $\M_{mn}(z)$ corresponding to the pinching of a pole in $F_{mn}(c(z))$ at $c_{i_mr_1}=0$, as in (\ref{rpoles}(c)),  with the contour in (\ref{ointhFz2}). $F_{mn}(c(z))$ has such a pole provided that $p>1$, that is in all cases except for split-helicity amplitudes. The potential contributions of this form come from the solutions of the equations (\ref{caseC})
\begin{align}
c_{i_ar_b}=0, \quad m'< a\leq m,\; 1\leq b\leq n' ;\qquad &c^{i_{a}i_{a+1}}_{r_{b}r_{b+1}}(z)=0,
\quad 1\leq a < m',\; n'<b<n;\cr
\C_{ab}=0, \quad 2\leq a\leq m'-1,\;2\leq b\leq n'\quad\hbox{or}&\quad m'< a\leq m-1,\;n'+2\leq b\leq n-1.&
\nonumber
\end{align}
and are labeled by integers $(m',n')$, $2\leq m'\leq m-1,\; 1\leq n'\leq n-2$. We shall investigate for which values of $(m',n')$ these correspond to nonzero solutions, showing that this happens precisely when $(m',n')$ lies on the `snake' line, and that then it gives the appropriate contribution to the BCFW relation (\ref{BCFWE}).

To evaluate the potential contributions $(m',n')$ specified by (\ref{caseC}), it is convenient to rewrite (\ref{ResM}),
 \be \hbox{Res}_{z_{m'n'}}\M_{mn}=K_{mn}\oint_{\overline\Oc} F_{mn}(c(z)) \prod_{a=2}^{m-1} \prod_{b=2}^{n-1} 
{d c_{i_ar_b}\over \C_{ab}( c(z))}dz,\label{ResMC}\ee
where now the contour $\overline{\Oc}$ encircles  $c_{i_mr_1}=0$ and the zeros of the constraints $\C_{ab}$, by rewriting some of these constraints. The set of constraints 
\be
\C_{ab}\equiv\C^{i_{a-1}i_ai_{a+1}}_{r_{b-1}r_br_{b+1}}=0,\; c\leq a\leq m-1,\; 2\leq b\leq d,
\ee
is equivalent to the set
\be
\C^{i_{a-1}i_ai_{a+1}}_{r_{b-1}r_br_{b+1}}=0,\; c\leq a\leq m-1,\; 2\leq b\leq d,\;(a,b)\ne(c,d);\quad\C^{i_{c-1}i_{c}i_{m}}_{r_{1}r_{d}r_{d+1}}=0,
\ee
so that, in effect, we have replaced the constraint $\C_{cd}\equiv\C^{i_{c-1}i_{c}i_{c+1}}_{r_{d-1}r_{d}r_{d+1}}=0$ with $\widehat\C_{cd}\equiv\C^{i_{c-1}i_{c}i_{m}}_{r_{1}r_{d}r_{d+1}}=0$. Using the techniques section of 2.1 of \cite{DGsh}, we can calculate the Jacobian factor that needs to be included if we make this replacement in (\ref{ResMC}),
\be 
\prod_{a=c}^{m-1} \prod_{b=2}^{d} {1\over \C_{ab}}\rightarrow\left[{c^{i_ci_m}_{r_1r_d}c_{i_{c-1}r_1}c_{i_mr_{d+1}}\over c^{i_ci_{c+1}}_{r_{d-1}r_d}c_{i_{c-1}r_{d-1}}c_{i_{c+1}r_{d+1}}}\right]
{1\over \widehat\C_{cd}}
\mathop{\prod\prod}\limits^{\scriptscriptstyle m-1\;\;\; d\;}_{a=c\;\; b=2\atop (a,b)\ne(c,d)}
{1\over \C_{ab}}\label{shift}
\ee
For the purpose of evaluating the potential contribution $(m',n')$ to (\ref{ResMC}), we  shall divide the constraints into four sets as in the products:
\be 
\prod_{a=2}^{m-1} \prod_{b=2}^{n-1} {1\over \C_{ab}}=\prod_{a=2}^{m'-1} \prod_{b=2}^{n'} {1\over \C_{ab}}\prod_{a=m'+1}^{m-1} \prod_{b=n'+2}^{n-1} {1\over \C_{ab}}
\mathop{\prod\prod}\limits^{\scriptscriptstyle m-1\;n'+1\;}_{a=m'\;\; b=2\atop (a,b)\ne(m',n'+1)}
\hskip-9pt{1\over \C_{ab}}\;\;\prod_{a=2}^{m'} \prod_{b=n'+1}^{n-1} {1\over \C_{ab}}.
\ee
The first two products will provide the constraints for subamplitudes $\M_{m',n'+1}$, $\M_{m''+1,n''}$, when the contribution is nonzero; the constraints in the last two products we shall rewrite and simplify. The fourth product can be rewritten using (\ref{shift}), starting with  $\C_{2,n-1}$ and working towards $\C_{m',n'+1}$,
\be 
\prod_{a=2}^{m'} \prod_{b=n'+1}^{n-1} {1\over \C_{ab}}\rightarrow\prod_{a=2}^{m'} \prod_{b=n'+1}^{n-1} 
\left[{c^{i_ai_m}_{r_1r_b}c_{i_{a-1}r_1}c_{i_mr_{b+1}}\over c^{i_ai_{a+1}}_{r_{b-1}r_b}c_{i_{a-1}r_{b-1}}c_{i_{a+1}r_{b+1}}}\right]
{1\over \widehat\C_{ab}}.
\label{shift2}
\ee
As $c_{i_mr_1}\rightarrow 0$, $\widehat\C_{ab}\rightarrow 
-c_{i_{m}r_{b}}c_{i_{m}r_{b+1}}c_{i_{a-1}r_{1}}c_{i_{a}r_{1}}c^{i_{a-1}i_{a}}_{r_{b}r_{b+1}},$
$\;c^{i_ai_m}_{r_1r_b}\rightarrow c_{i_{a}r_{1}}c_{i_{m}r_{b}}$. In this way, we can successively replace the condition
$\C_{ab}=0$ by $c^{i_{a-1}i_{a}}_{r_{b}r_{b+1}}=0$, the factors of $ c^{i_ai_{a+1}}_{r_{b-1}r_b}$ in (\ref{shift2}) canceling those in the numerator of  $F^{\hbox{\tiny split}}_{mn}(c)$ and so in the numerator of 
$F_{mn}(c)=G_{mn}(c)F^{\hbox{\tiny split}}_{mn}(c)$, unless $G_{mn}(c)$ provides extra factors. Thus,
\be 
\prod_{a=2}^{m'} \prod_{b=n'+1}^{n-1} {1\over \C_{ab}}\rightarrow\prod_{a=2}^{m'} \prod_{b=n'+1}^{n-1} 
\left[-{1\over c^{i_ai_{a+1}}_{r_{b-1}r_b}c_{i_{a-1}r_{b-1}}c_{i_{a+1}r_{b+1}}}\right]
{1\over c^{i_{a-1}i_{a}}_{r_{b}r_{b+1}}}.
\label{shift3}
\ee

Using 
\be 
\C_{ab}=-c_{i_{a-1}r_{b-1}}c_{i_{a}r_{b-1}}c_{i_{a+1}r_{b}}c_{i_{a+1}r_{b+1}} c^{i_{a-1}i_a}_{r_br_{b+1}} \quad \hbox{if}\quad c_{i_{a+1}r_{b-1}}=0,
\ee
gives
 \begin{align}
  \hbox{Res}_{z_{m'n'}}\M_{mn}=K_{mn}&\oint_{\widetilde\Oc}  F^{\hbox{\tiny split}}_{\sss m',n'+1}\prod_{\sss a=2}^{\sss m'-1} \prod_{\sss b=2}^{n'} {1\over \C_{ab}}F^{\hbox{\tiny split}}_{\sss m''+1,n''}\prod_{\sss a=m'+1}^{\sss m-1} \prod_{\sss b=n'+2}^{\sss n-1} {1\over \C_{ab}}\left[-{c_{i_mr_1}c_{i_{m'}r_{n'+1}}^2\over c_{i_{m'}r_1}c_{i_{m}r_{n'+1}}}\right]G_{mn}\cr
  &\times \prod_{a=2}^{m'} \prod_{b=n'+1}^{n-1} \left[-{c_{i_ar_b}\over c^{i_{a-1}i_{a}}_{r_{b}r_{b+1}}} \right]\prod_{a=m'+1}^{m} \prod_{b=1}^{n'}  \left[-{1\over c_{i_ar_b}}\right]
 \prod_{a=2}^{m-1} \prod_{b=2}^{n-1} 
d c_{i_ar_b}dz,\label{ResMC2}\end{align}
where the contour $\widetilde\Oc$ encircles the solution of (\ref{caseC}) and $F^{\hbox{\tiny split}}_{\sss m',n'+1}$ and 
$F^{\hbox{\tiny split}}_{\sss m''+1,n''}$
have arguments $c_{i_ar_b}$ for $1\leq a\leq m', 1\leq b\leq n'+1$ and for $m'\leq a\leq m, n'+1\leq b\leq n,$ respectively.

The expression (\ref{ResMC2}) for the residue at $z=z_{m'n'}$ will vanish if $G_{mn}$, defined by (\ref{defG}), contains a factor in its numerator which is zero on the set defined by (\ref{caseC}) at which we are calculating the residue. Thus none of the factors $c_{i_{m_e}r_{n_{e-1}+1}}$, $1\leq e\leq p$, must be among the set $c_{i_ar_b}$, $m'+1\leq a\leq m, 1\leq b\leq n'$, so that we must have 
\be
m'\geq m_e,  \quad\hbox{if}\quad  n'>n_{e-1}, \qquad 1\leq e\leq p,\label{s1}
\ee
and
none of the factors $c^{i_{m_e}i_{m_e+1}}_{r_{n_e}r_{n_e+1}}$, $1\leq e\leq p-1$, must be among the set $c^{i_{a}i_{a+1}}_{r_{b}r_{b+1}}$, $1\leq a\leq m'-1, n'+1\leq b\leq n-1$, so that 
\be 
m' \leq m_e\quad\hbox{if}\quad n'< n_e, \qquad 1\leq e\leq p-1.\label{s2}
\ee
From (\ref{s1}) and (\ref{s2}), it follows that 
\be
m'=m_e\; \hbox{ if } \; n_{e-1}< n'<n_{e},\quad\hbox{and}\quad
\quad m_{e}\leq m'\leq m_{e+1}\; \hbox{ if } \; n'=n_{e},\;\; 1\leq e\leq p-1.
\label{s3}\ee
Thus we see that the contribution (C) will vanish unless $(m',n')$ lies on the `snake' line. 

If $(m',n')$ lies on the `snake' line, we can form the corresponding lines for the associated amplitudes $\M_{m',n'+1}$ and $ \M_{m''+1,n''}$ that occur in the negative helicity contribution to the BCFW relation by severing the line in two at this point and adding a horizontal step to the end of the first part and vertical step to the beginning of the second part. If $G_{ m',n'+1}^L$ and $G_{ m''+1,n''}^R$ denote the functions associated, as in (\ref{GF}), with the two subamplitudes, they satisfy the following relation to $G_{mn}$, 
\be
\left[{c_{i_mr_1}c_{i_{m'}r_{n'+1}}\over c_{i_{m'}r_1}c_{i_{m}r_{n'+1}}}\right]G_{\sss mn}
=G_{\sss m',n'+1}^LG_{\sss m''+1,n''}^R,\label{sG}
\ee
provided that $c_{i_{m'+1}r_{n'}}=0$. To establish (\ref{sG}), consider the possibilities for the position of $(m',n')$ on the snake line: either it is on a vertical part of the line, or it is on an horizontal part, or it is at a corner of the form $(m_e, n_{e})$, or at a corner of the form $(m_e, n_{e-1})$. In each of the first three cases, there is one more corner of the first form for the two `snake' lines for the subamplitudes taken together than for that for the original amplitude and this produces a factor of $c_{i_{m'}r_{n'+1}}$; in the fourth case, there are no corners at $(m',n')$ on the `snake' lines for the subamplitudes, but the factor associated  with the corner of the second form reduces to the inverse of 
$c_{i_{m'}r_{n'+1}}$ when $c_{i_{m'+1}r_{n'}}=0$. The remaining factors on the left hand side of (\ref{sG}) correspond to the ratio of the first factors in the definition (\ref{defG}) for the amplitude and subamplitudes, thus establishing the relation (see also Appendix \ref{RelG}). 

Performing the integrations with respect to $c_{i_ar_b}$, $m'\leq a\leq m-1, 2\leq b\leq n'+1$, and $2\leq a\leq m', n'+1\leq b\leq n-1$, and $z$, and using the expression for the relevant Jacobian given in (\ref{detC}),
\begin{align}
  \hbox{Res}_{z_{m'n'}}\M_{mn}=&-{K_{m',n'+1}K_{m''+1,n''}\over[i_1|P_{m'n'}|r_n\rangle}
  \oint_{\Oc_L} \mu_{n'+1}^{m'} F^{\hbox{\tiny split}}_{m',n'+1}G_{\sss m',n'+1}^L\prod_{a=2}^{m'-1} \prod_{b=2}^{n'} {dc_{i_ar_b}\over \C_{ab}}\cr
  &\hskip7truemm \times \oint_{\Oc_R}\lambda_{m'}^{n''}F^{\hbox{\tiny split}}_{m''+1,n''}G_{\sss m''+1,n''}^R\prod_{a=m'+1}^{m-1} \prod_{b=n'+2}^{n-1}{dc_{i_ar_b}\over \C_{ab}},
  \label{ResMC3}\end{align}
  where the contours $\Oc_L$ and $\Oc_R$ encircle the solutions of $\C_{ab}=0$ for $2\leq a\leq m'-1,\;2\leq b\leq n'$, $m'+1\leq a\leq m-1, \; n'+2\leq b\leq n-1$, respectively,  
  \be
  K_{m',n'+1}=\langle r_{1},r_{0}\rangle^{m'-2}[i_1,i_{m'}]^{n'-1},\quad K_{m''+1,n''}=[i_{0},i_m]^{n''-2}\langle r_{n'+1},r_{n}\rangle^{m''-1},
   \ee
   and $c_{i_ar_b}=\lambda_a\mu_b$ for $1\leq a\leq m',\; n'<b\leq n,\;(a,b)\ne (1,n),$ as in (\ref{clmC}).

Defining 
\begin{alignat}{3}
c^L_{i_ar_b}&= c_{i_ar_b},&\quad c_{i_ar_{n'+1}}^L&= \lambda_a,\quad 1\leq a\leq m',\;1\leq b\leq n',\cr
c^R_{i_ar_b}&= c_{i_ar_b},&\quad c_{i_{m'}r_b}^R&= \mu_b,\quad m'< a\leq m,\;n'< b\leq n,\cr
\end{alignat}
and $F_{m',n'+1}=F^{\hbox{\tiny split}}_{m',n'+1}G_{m',n'+1}^L$, \;$F_{m''+1,n''}=F^{\hbox{\tiny split}}_{m''+1,n''}G_{m''+1,n''}^R,$
\be
F_{m',n'+1}(c)=\mu_{n'+1}^{m'-4}F_{m',n'+1}(c^L), \quad F_{m''+1,n''}(c)=\lambda_{m'}^{n''-4}F_{m''+1,n''}(c^R)
\ee
so that, using (\ref{sG}) and (\ref{zPmn}), we can rewrite (\ref{ResMC3}) in the form
\be
\hbox{Res}_{z_{m'n'}}\M_{mn}=-{z_{m'n'}\over s_{m'n'}}\M_{m',n'+1}\M_{m''+1,n''}\label{contC}
\ee
where the arguments of $\M_{m',n'+1}$ and $\M_{m''+1,n''}$ are the momenta described in (\ref{mb}) and (\ref{mc}), and we obtain the appropriate contribution to BCFW relation (\ref{BCFWE}) whenever the point $(m',n')$ lies on the `snake' line described in section \ref{FormF}.

\section{\bf Contributions of Type (D) }\label{TypeD}
\nobreak
Lastly, we consider contributions to  (\ref{exppole}) arising from poles in $\M_{mn}(z)$ corresponding to the pinching of a pole in $F_{mn}(c(z))$ at $c_{i_1r_{n-1}}=0$, as in (\ref{spoles}(d)),  $c_{i_2r_{n}}=0$, as in (\ref{spoles}(e)),  or $c_{i_1r_{n}}=0$, as in (\ref{spoles}(f)),  with the contour in (\ref{ointhFz2}).  The potential contributions of this form come from the solutions of the equations (\ref{caseD})
\begin{align}
  c_{i_ar_b}(z)=0, \quad 1\leq a\leq m',\; n'< b\leq n ;\qquad &c^{i_{a}i_{a+1}}_{r_{b}r_{b+1}}=0,
\quad m'< a < m,\; 1\leq b<n';\cr
\C_{ab}=0, \quad 2\leq a\leq m',\;2\leq b\leq n'-1\quad\hbox{or}&\quad m'< a\leq m-1,\;n'+1\leq b\leq n-1.&
\label{caseD2}
\end{align}
and are 
labeled by integers $(m',n')$, $1\leq m'\leq m-2,\; 2\leq n'\leq n-1$, or $(m',n')=(1,1)$ or $(m-1,n-1)$. The equations (\ref{caseD2}) also describe `spurious' contributions \cite{DG3}, which occur when there are contributions to (\ref{ointhF}) corresponding to points at which some $c_{js}=0$ which need to be subtracted. These `spurious' contributions and those corresponding to poles of the form (\ref{spoles}) have the same form. We shall investigate for which values of $(m',n')$ these correspond to nonzero solutions, showing again that this happens precisely when $(m',n')$ lies on the `snake' line, and that then it gives the appropriate contribution to the BCFW relation (\ref{BCFWE}).

We follow the approach of section \ref{TypeC} to evaluate the potential contributions $(m',n')$ specified by (\ref{caseD}), and rewrite (\ref{ResM}),
 \be \hbox{Res}_{z_{m'n'}}\M_{mn}=K_{mn}\oint_{\overline\Oc} F_{mn}(c(z)) \prod_{a=2}^{m-1} \prod_{b=2}^{n-1} 
{d c_{i_ar_b}\over \C_{ab}( c(z))}dz,\label{ResMD}\ee
where now the contour $\overline{\Oc}$ encircles  $c_{i_1r_n}=0$ and the zeros of the constraints $\C_{ab}$, and we shall again rewrite some of these constraints. The set of constraints 
\be
\C_{ab}\equiv\C^{i_{a-1}i_ai_{a+1}}_{r_{b-1}r_br_{b+1}}=0,\; 2\leq a\leq c,\; d\leq b\leq n-1,
\ee
is equivalent to the set
\be
\C^{i_{a-1}i_ai_{a+1}}_{r_{b-1}r_br_{b+1}}=0,\; 2\leq a\leq c,\; d\leq b\leq n-1,\;(a,b)\ne(c,d);\quad\C^{i_1i_{c}i_{c+1}}_{r_{d-1}r_{d}r_{n}}=0,
\ee
where the constraint $\C_{cd}\equiv\C^{i_{c-1}i_{c}i_{c+1}}_{r_{d-1}r_{d}r_{d+1}}=0$ has been replaced with $\widetilde\C_{cd}\equiv\C^{i_1i_{c}i_{c+1}}_{r_{d-1}r_{d}r_{n}}=0$.
Following (\ref{shift}), we can make the replacement 
 \be 
\prod_{a=2}^{c} \prod_{b=d}^{n-1} {1\over \C_{ab}}\rightarrow
\left[{c^{i_1i_c}_{r_dr_n}c_{i_{c+1}r_n}c_{i_1r_{d-1}}\over c^{i_{c-1}i_{c}}_{r_{d}r_{d+1}}c_{i_{c-1}r_{d-1}}c_{i_{c+1}r_{d+1}}}\right]
{1\over \widetilde\C_{cd}}
\mathop{\prod\prod}\limits^{\scriptscriptstyle \;\;\;c\;\;\;\; n-1\;}_{a=2\;\; b=d\atop (a,b)\ne(c,d)}
{1\over \C_{ab}}\label{shift4}
\ee
in (\ref{ResMD}).
To evaluate the potential contribution $(m',n')$ to (\ref{ResMD}), we  again divide the constraints into four sets as in the products:
\be 
\prod_{a=2}^{m-1} \prod_{b=2}^{n-1} {1\over \C_{ab}}=\prod_{a=2}^{m'} \prod_{b=2}^{n'-1} {1\over \C_{ab}}\prod_{a=m'+2}^{m-1} \prod_{b=n'+1}^{n-1} {1\over \C_{ab}}
\mathop{\prod\prod}\limits^{\scriptscriptstyle m'+1\;n-1\;}_{a=2\;\; b=n'\atop (a,b)\ne(m'+1,n')}
\hskip-9pt{1\over \C_{ab}}\;\;\prod_{a=m'+1}^{m-1} \prod_{b=2}^{n'} {1\over \C_{ab}}.
\ee
The first two products will provide the constraints for subamplitudes $\M_{m'+1,n'}$, $\M_{m'',n''+1}$, when the contribution is nonzero; we shall rewrite the constraints in the last two products. To rewrite the fourth product, we use (\ref{shift4}), starting with  $\C_{m-1,2}$ and working towards $\C_{m'+1,n'}$,
\be 
\prod_{a=m'+1}^{m-1} \prod_{b=2}^{n'} {1\over \C_{ab}}\rightarrow
\prod_{a=m'+1}^{m-1} \prod_{b=2}^{n'}
\left[{c^{i_1i_a}_{r_br_n}c_{i_{a+1}r_n}c_{i_1r_{b-1}}\over c^{i_{a-1}i_{a}}_{r_{b}r_{b+1}}c_{i_{a-1}r_{b-1}}c_{i_{a+1}r_{b+1}}}\right]
{1\over \widetilde\C_{ab}}.
\label{shift5}
\ee

As $c_{i_1r_n}\rightarrow 0$, $\widetilde\C_{ab}\rightarrow 
-c_{i_{1}r_{b-1}}c_{i_{1}r_{b}}c_{i_{a}r_{n}}c_{i_{a+1}r_{n}}c^{i_{a}i_{a+1}}_{r_{b-1}r_{b}},$
$\;c^{i_1i_a}_{r_br_n}\rightarrow c_{i_{1}r_{b}}c_{i_{a}r_{n}}$. In this way, we can successively replace the condition
$\C_{ab}=0$ by $c^{i_{a}i_{a+1}}_{r_{b-1}r_{b}}=0$, the factors of $c^{i_{a-1}i_{a}}_{r_{b}r_{b+1}}$ in (\ref{shift4}) canceling those in the numerator of  $F^{\hbox{\tiny split}}_{mn}(c)$ and so in the numerator of 
$F_{mn}(c)=G_{mn}(c)F^{\hbox{\tiny split}}_{mn}(c)$, unless $G_{mn}(c)$ provides extra factors. Thus,
\be 
\prod_{a=m'+1}^{m-1} \prod_{b=2}^{n'} {1\over \C_{ab}}\rightarrow
\prod_{a=m'+1}^{m-1} \prod_{b=2}^{n'}
\left[-{1\over c^{i_{a-1}i_{a}}_{r_{b}r_{b+1}}c_{i_{a-1}r_{b-1}}c_{i_{a+1}r_{b+1}}}\right]
{1\over c^{i_{a}i_{a+1}}_{r_{b-1}r_{b}}}.
\label{shift6}
\ee

Using 
\be 
\C_{ab}=-c_{i_{a-1}r_{b-1}}c_{i_{a-1}r_{b}}c_{i_{a}r_{b+1}}c_{i_{a+1}r_{b+1}} c^{i_{a}i_{a+1}}_{r_{b-1}r_{b}} \quad \hbox{if}\quad c_{i_{a-1}r_{b+1}}=0,
\ee
gives
\begin{align}
  \hbox{Res}_{z_{m'n'}}\M_{mn}=K_{mn}&\oint_{\widetilde\Oc}  F^{\hbox{\tiny split}}_{\sss m'+1,n'}\prod_{\sss a=2}^{\sss m'} \prod_{\sss b=2}^{n'-1} {1\over \C_{ab}}F^{\hbox{\tiny split}}_{\sss m'',n''+1}\prod_{\sss a=m'+2}^{\sss m-1} \prod_{\sss b=n'+1}^{\sss n-1} {1\over \C_{ab}}
 \left[-{c_{i_1r_n}c_{i_{m'+1}r_{n'}}^2\over c_{i_1r_{n'}}c_{i_{m'+1}r_n}}\right]G_{mn}\cr
  &\times \prod_{a=m'+1}^{m-1} \prod_{b=2}^{n'} \left[-{c_{i_ar_b}\over c^{i_{a}i_{a+1}}_{r_{b-1}r_{b}}} \right]
 \prod_{a=1}^{m'}\prod_{b=n'+1}^{n}   \left[-{1\over c_{i_ar_b}}\right]
 \prod_{a=2}^{m-1} \prod_{b=2}^{n-1} 
d c_{i_ar_b}dz,\label{ResMD2}\end{align} 
where the contour $\widetilde\Oc$ encircles the solution of (\ref{caseD}) and $F^{\hbox{\tiny split}}_{\sss m'+1,n'}$ and 
$F^{\hbox{\tiny split}}_{\sss m'',n''+1}$
have arguments $c_{i_ar_b}$ for $1\leq a\leq m'+1, \;1\leq b\leq n'$ and for $m'+1\leq a\leq m, \;n'\leq b\leq n,$ respectively. It is convenient to use for $G_{mn}$ the second form (\ref{Gpn2}), equivalent to (\ref{defG}), given in Appendix \ref{RelG},
\be
G_{mn}={ c_{i_1r_1}c_{i_1r_n}c_{i_mr_n}\over c^{i_1i_m}_{r_1r_n}}
\prod_{e=1}^{p-1}c_{i_{m_e+1}r_{n_{e}}}\prod_{e=1}^{p}{c^{i_{m_e}i_{m_e+1}}_{r_{n_{e-1}}r_{n_{e-1}+1}}
\over c_{i_{m_e}r_{n_{e-1}}}c_{i_{m_e+1}r_{n_{e-1}}}c_{i_{m_e+1}r_{n_{e-1}+1}}}.\label{Galt}
\ee

The expression (\ref{ResMD2}) for the residue at $z=z_{m'n'}$ will vanish if $G_{mn}$, defined by (\ref{defG}), contains a factor in its numerator which is zero on the set defined by (\ref{caseD}) at which we are calculating the residue. Thus none of the factors $c_{i_{m_e+1}r_{n_{e}}}$, $1\leq e\leq p$, must be among the set $c_{i_ar_b}$, $1\leq a\leq m',\; n'+1\leq b\leq n $, so that we must have 
\be 
m' \leq m_e\quad\hbox{if}\quad n'< n_e, \qquad 1\leq e\leq p-1.\label{t1}
\ee
and, using the form (\ref{Gpn2}) for $G_{mn}$,
none of the factors $c^{i_{m_1}i_{m_1+1}}_{r_{1}r_{2}}$, $c^{i_{m_e}i_{m_e+1}}_{r_{n_{e-1}}r_{n_{e-1}+1}}$, $2\leq e\leq p-1$, $c^{i_{m-1}i_{m}}_{r_{n_{p-1}}r_{n_{p-1}+1}}$, must be among the set $c^{i_{a}i_{a+1}}_{r_{b}r_{b+1}}$, $m'+1 \leq a\leq m-1, 1\leq b\leq n'-1$, so that 
\be
m'\geq m_e,  \quad\hbox{if}\quad  n'>n_{e-1}, \qquad 1\leq e\leq p.\label{t2}
\ee
Conditions (\ref{t1}) and (\ref{t2}) are the same as (\ref{s2}) and (\ref{s1}), respectively, and so it follows that 
\be
m'=m_e\; \hbox{ if } \; n_{e-1}< n'<n_{e},\quad\hbox{and}\quad
\quad m_{e}\leq m'\leq m_{e+1}\; \hbox{ if } \; n'=n_{e},\;\; 1\leq e\leq p-1,
\label{t3}\ee
implying that the contribution (D) will vanish unless $(m',n')$ lies on the `snake' line.

If $(m',n')$ lies on the `snake' line, we form the corresponding lines for the associated amplitudes $\M_{m'+1,n'}$ and $ \M_{m'',n''+1}$ that occur in the positive helicity contribution to the BCFW relation by again severing the line at this point and adding a vertical step to the end of the first part and a horizontal step to the beginning of the second part. If $G_{ m'+1,n'}^L$ and $G_{ m'',n''+1}^R$ denote the functions associated, as in (\ref{GF}), with the two subamplitudes, from  (\ref{pGrel}), they satisfy the following relation to $G_{mn}$, 
\be
\left[{c_{i_1r_n}c_{i_{m'+1}r_{n'}}\over c_{i_1r_{n'}}c_{i_{m'+1}r_n}}\right]
G_{\sss mn}
=G_{\sss m'+1,n'}^{L}G_{\sss m'',n''+1}^{R},\label{pG}
\ee
provided that $c_{i_1r_n}=c_{i_{m'}r_{n'+1}}=c^{i_mi_{m'+1}}_{r_1r_{n'}} =0$, which hold for contributions of type D.

Performing the integrations with respect to $c_{i_ar_b}$, $m'+1\leq a\leq m-1, 2\leq b\leq n'$, and $2\leq a\leq m'+1, n'\leq b\leq n-1$, and $z$, and using the expression for the relevant Jacobian given in (\ref{detD}),
\begin{align}
  \hbox{Res}_{z_{m'n'}}\M_{mn}=&-{K_{m'+1,n'}K_{m'',n''+1}\over[i_1|P_{m'n'}|r_n\rangle}
  \oint_{\Oc_L} \lambda_{m'+1}^{n'} F^{\hbox{\tiny split}}_{m'+1,n'}G_{\sss m'+1,n'}^L\prod_{a=2}^{m'} \prod_{b=2}^{n'-1} {dc_{i_ar_b}\over \C_{ab}}\cr
  &\hskip7truemm \times \oint_{\Oc_R}\mu_{n'}^{m''}F^{\hbox{\tiny split}}_{m'',n''+1}G_{\sss m'',n''+1}^R\prod_{a=m'+2}^{m-1} \prod_{b=n'+1}^{n-1}{dc_{i_ar_b}\over \C_{ab}},
  \label{ResMD3}\end{align}
  where the contours $\Oc_L$ and $\Oc_R$ encircle the solutions of $\C_{ab}=0$ for $2\leq a\leq m',\;2\leq b\leq n'-1$, $m'+2\leq a\leq m-1, \; n'+1\leq b\leq n-1$, respectively,  
 \be
  K_{m'+1,n'}=\langle r_{1},r_{n'}\rangle^{m'-1}[i_1,i_{0}]^{n'-2},\quad K_{m'',n''+1}=[i_{m'+1},i_m]^{n''-1}\langle r_{0},r_{n}\rangle^{m''-2},
   \ee
   and $c_{i_ar_b}=\lambda_a\mu_b$ for $m'< a\leq m,\;1\leq b\leq n',\;(a,b)\ne (1,n),$ as in (\ref{clmD}).

Defining 
\begin{alignat}{3}
c^L_{i_ar_b}&= c_{i_ar_b},&\quad c_{i_{m'+1}r_{b}}^L&= \mu_b,\quad 1\leq a\leq m',\;1\leq b\leq n',\cr
c^R_{i_ar_b}&= c_{i_ar_b},&\quad c_{i_{a}r_{n'}}^R&= \lambda_a,\quad m'< a\leq m,\;n'< b\leq n,
\end{alignat}
and $F_{m'+1,n'}=F^{\hbox{\tiny split}}_{m'+1,n'}G_{m'+1,n'}^L$, \;$F_{m'',n''+1}=F^{\hbox{\tiny split}}_{m'',n''+1}G_{m'',n''+1}^R,$
\be
F_{m'+1,n'}(c)=\lambda_{m'+1}^{n'-4}F_{m'+1,n'}(c^L), \quad F_{m'',n''+1}(c)=\mu_{n'}^{m''-4}F_{m'',n''+1}(c^R)
\ee
so that (\ref{ResMD3}) can be written
\be
\hbox{Res}_{z_{m'n'}}\M_{mn}=-{z_{m'n'}\over s_{m'n'}}\M_{m'+1,n'}\M_{m'',n''+1}\label{contD}
\ee
where the arguments of $\M_{m'+1,n'}$ and $\M_{m'',n''+1}$ are the momenta described in (\ref{md}) and (\ref{me}), and we obtain the appropriate contribution to BCFW relation (\ref{BCFWE}) whenever the point $(m',n')$ lies on the `snake' line. Taking this together with the results of sections \ref{TypeAB} and \ref{TypeC}, we have established the BCFW relation provided that there are no contributions from other poles of $\M_{mn}(z)$. We shall establish this in the next section.

\section{\bf Absence of Other Contributions}
\label{Abs}

As we discussed in section \ref{PolesM},  the poles of $M_{mn}(z)$, are generated by pinches between poles in the integrand corresponding to factors in the denominator of $F_{mn}$ and poles corresponding to the constraints $\C_{ab}$, and these are determined by the intersection between the hyperplane specified by the momentum constraints (\ref{L1}) and  surfaces of solutions to (\ref{csol}) of dimension $2(m+n)-4$, that is codimension $(m-2)(n-2)+1$ in the space of the $c_{i_ar_b}, 1\leq a\leq m, 1\leq b\leq n$, and $z$.. In this section, we analyze the possible form of such surfaces. 

The possible poles in $F_{mn}$ are listed in (\ref{rpoles}) and (\ref{spoles}). Those arising from factors $c^{i_1i_2}_{r_1r_2}$ (\ref{rpoles}a) and $c^{i_{m-1}i_m}_{r_{n-1}r_n}$ (\ref{rpoles}b) were discussed in Section \ref{TypeAB} and lead to the ($2(m+n)-4$)-dimensional surfaces (\ref{caseA}) and (\ref{caseB}), respectively. In other cases, the poles in $F_{mn}$ correspond to factors $c_{ir}$ for some particular $i$ and $r$. We now discuss this case. Consider the surface of codimension $(m-2)(n-2)+1$ specified by the conditions 
\begin{alignat}{5}
&(i) &\quad\C^{jkl}_{rst}&=0 &&\qquad\hbox{for } j,k,l\in I,\; s,t,u\in R;\cr
&(ii)&\quad c_{ir}&=0&&\qquad\hbox{for a particular } i\in I\hbox{ and } r\in R,\label{hyp}
\end{alignat}
where $|I|=m$ and $|R|=n$. We shall show by induction on $m$ and $n$ that the space of solutions comprises surfaces of the form
\begin{alignat}{5}
&(a)&\quad c_{js}&=0&&\qquad j\in I_2,\; s\in R_1,\cr
&(b)&\quad c^{jk}_{st}&=0&&\qquad  j,k\in {I}_1, \;  s,t\in R_2,\cr
&(c) &\quad \C^{jkl}_{stu}&=0 &&\qquad\hbox{for } j,k,l\in I_1,\; s,t,u\in \overline{R}_1;\cr
&(d) &\quad \C^{jkl}_{stu}&=0 &&\qquad\hbox{for } j,k,l\in \overline I_2,\; s,t,u\in R_2,\label{thm}
\end{alignat}
where $I_1,I_2$ are complementary subsets in $I$, {\it i.e.} $I=I_1\cup I_2$, $I_1\cap I_2=\O$, and similarly for $R_1, R_2$ in $R$; $ \overline I_2=I_2\cup\{i'\}$, $ \overline R_1=R_1\cup\{r'\}$, where 
$i'\in I_1,\; r'\in R_2$ are arbitrarily chosen. The choice does not matter because of (b). 

The result is easily seen to be true for $m=n=3$, because, from (\ref{C''}), if $c_{ir}=0$, $\C^{ijk}_{rst}=-c_{is}c_{it}c_{jr}c_{kr}c^{jk}_{st}$ and one of these five factors has to vanish. If $c^{jk}_{st}=0$, we have (\ref{thm}) with $I_2=\{i\}, R_1=\{r\}$; if $c_{is}=0$, we have (\ref{thm}) with $I_2=\{i\}, R_1=\{r,s\}$, and similarly for the other factors.

Now suppose the result holds for particular values of $m$ and $n$, and suppose we seek to prove it for $m+1$, $n$, with $I$ replaced by $I'=I\cup\{i_\ast\}$. We assume the hypothesis (\ref{hyp}) holds for $I', R$. Then consider $\C^{jki_\ast}_{stu}=0$, where $j\in I_2, k\in I_1$, $s\in R_1, t,u\in R_2$. Then $c_{jt}c_{ju}c_{ks}c_{i_\ast s}c^{ki_\ast}_{tu}=0,$ then $c_{jt},c_{ju},c_{ks}$ are each generically nonzero, or we should have too many conditions on the surface, so that either $c_{i_\ast s}=0$ or $c^{ki_\ast}_{tu}=0$. Since $s\in R_1$ is arbitrary,  we have that $c^{ki_\ast}_{tu}=0$, unless $c_{i_\ast s}=0$ for all $s\in R_1$.  In the latter case,  (\ref{thm}) holds with $I_2$ replaced by $I_2'=I_2\cup \{i_\ast\}$. Otherwise, $c^{ki_\ast}_{tu}=0$ for all $k\in I_1$, $t,u\in R_2$ and  (\ref{thm}) holds with $I_1$ replaced by $I_1'=I_1\cup \{i_\ast\}$, establishing the induction and hence the result. 

The conditions (\ref{thm}), together with the momentum constraints (\ref{L1z}), imply an equation of the form of (\ref{momc}),
\be
\sum_{i \in I_1}\bpi_i\pi_i+\sum_{r\in R_1} \bpi_r\pi_r=
-\sum_{i \in I_2}\bpi_i\pi_i-\sum_{r\in R_2} \bpi_r\pi_r,
\label{momd}
\ee
where $\pi_{i_1}$ stands for $\pi_{i_1}(z)=\pi_{i_1}+z\pi_{r_n}$ and $\bpi_{r_n}$ stands for $\bpi_{r_n}(z)=\bpi_{r_n}-z\bpi_{i_1}$. If {\it either} $i_1\in I_1$ and $r_n\in R_1$ {\it or} $i_1\in I_2$ and $r_n\in R_2$, the $z$-dependence cancels out from (\ref{momc}) and the conditions  (\ref{thm}) imply a relationship between momenta and do not correspond to the contribution of a pole in $\M_{mn}$. In the other two cases, $i_1\in I_1$ and $r_n\in R_2$ {\it or} $i_1\in I_2$ and $r_n\in R_1$, we have, as in (\ref{zPA}), 
\be 
z=z_A\equiv s_A/[i_1|P_A|r_n\rangle,\label{zPA2}
\ee
where 
 $A=I_1\cup R_1$ if $i_1\in I_1$ and $A=I_2\cup R_2$ if $i_1\in I_2$. We shall show that these two cases, which we label (1) and (2), respectively, correspond to case (C) (\ref{caseC}) and to case (D) (\ref{caseD}), respectively.

Now consider a contribution to 
\be \hbox{Res}_{z_{A}}\M_{mn}=K_{mn}\oint_{\overline\Oc} F^{\hbox{\tiny split}}_{mn}\; G_{mn}\prod_{a=2}^{m-1} \prod_{b=2}^{n-1} {d c_{i_ar_b}\over \C_{ab}}dz,\label{ResMC4}\ee
for which $c_{ir}=0$ for some particular $i\in I$ and $r\in R$. Then, if $\{i,j,k\}$ and $\{r,s,t\}$ are each sets of adjacent indices, $\C^{ijk}_{rst}$ is one of the constraints in the denominator of the integrand in (\ref{ResMC4}) and it factorizes, $\C^{ijk}_{rst}=-c_{is}c_{it}c_{jr}c_{kr}c^{jk}_{st}$, so that one or more of these factors has to vanish for the contribution. Similarly, if $c^{jk}_{st}=0$ for the contribution, and, $\{j,k,l\}$ and $\{s,t,u\}$ are each sets of adjacent indices, the constraint $\C^{jkl}_{stu}$ factorizes, $\C^{jkl}_{stu}=-c_{js}c_{lu}c^{jk}_{tu}c^{kl}_{st}$, so that one or more of these factors has to vanish for the contribution. Progressing in this way, by factorizing constraints, we generate successively from the initial condition $c_{ir}=0$ a set of conditions of the form $c_{js}=0$ and $c^{kl}_{tu}=0$ that, together with the remaining constraints, which are of the form $\C^{i_{a-1}i_ai_{a+1}}_{r_{b-1}r_br_{b+1}}$, defining a ($2(m+n)-4$)-dimensional surface. This surface must therefore be of the form (\ref{thm}), but with the particular feature that the indices in each of the individual sets $I_1, I_2, R_1, R_2$ are consecutive. For $|I_1|, |I_2|, |R_1|, |R_2|\geq 3$, this is clear but if one of them is smaller it needs a more detailed argument. If they are consecutive, it follows that they must be as in  case (C) (\ref{caseC}) and to case (D) (\ref{caseD}), depending on whether we are in case (1) or case (2), respectively, of (\ref{momd}) and (\ref{zPA2}).

To see in detail that $I_1, I_2, R_1, R_2$ are consecutive, suppose that $|R_1|,|I_2|\geq 2$ and $|I_1|, |R_2|\geq 3$, then considering  $\C^{ijk}_{rst}$ for any $i,j,k\in I_1, r, s\in R_1, t\in R_2$ shows that $I_1$ must consist of consecutive indices, and similarly for $R_2$. Also, it follows that, for any $r,s\in R_1$, $\{r, s,t\}$ must be consecutive for some $t\in R_2$ and, given that $R_2$ has been shown to be consecutive, $r,s$ can not be separated by the indices of $R_2$. Thus the indices of $R_1$ are consecutive and similarly for $I_2$, establishing that, if $|R_1|,|I_2|\geq 2$ and $|I_1|, |R_2|\geq 3$, the individual sets $I_1, I_2, R_1, R_2$ are consecutive.

If $|R_1|,|I_2|\geq 2$, $|I_1|\geq 3$ and $ |R_2|=2$, the same argument shows that $I_1$ is consecutive. If $i_1\in I_1$, so that $r_n\in R_2$ as in case (1), then $i_2\in I_1$ and $i_m\in I_2$;  if we also had $r_1\in R_2$, we should have a condition $c^{i_1i_2}_{r_1r_n}=0$, where $r_1, r_n$  are neither adjacent nor next to adjacent. But all the conditions $c^{ij}_{rs}=0$ that come from factorizing a constraint $\C^{ijk}_{rst}=0$ with consecutive indices, involve indices separated by no more than other index. It follows that we have $r_1\in R_1$ and $i_m\in I_2$. The alternative is to have  $i_1\in I_2$ and $r_n\in R_1$, as in case (2). In the former case, the denominator of (\ref{ResMC6}) will not vanish as $c^{i_1i_2}_{r_br_n}\rightarrow 0$ unless $b=n-1$, so we will not get a nonzero residue unless both $r_{n-1}, r_n\in R_2$ as well as $i_1, i_2\in I_1$; in the latter case, it follows similarly that  $i_{m-1}, i_m\in I_1, r_{1}, r_2\in R_2$, and so, in each case, the sets $I_1, I_2, R_1, R_2$ are each consecutive. The result follows similarly if $|R_1|,|I_2|\geq 2$, $|I_1|=2$ and $ |R_2|\geq 3$. If $|R_1|,|I_2|\geq 2$ and $|I_1|= |R_2|=2$, because the sole condition of the form $c^{ij}_{rs}=0$ has indices which are consecutive or next to consecutive, we can not have both $i_1\in I_1$ and $i_m\in I_1$ and we can not have both $r_1\in R_2$ and $r_n\in R_2$, so that if $i_1\in I_1$ and $r_n\in R_2$, as in case (1), then $i_m\in I_2$ and $r_1\in R_1$ and again, using the argument below, we have that the sets $I_1, R_2$ are again positioned in a corner and each of the sets is consecutive. Thus the result holds whenever $|I_1|, |I_2|, |R_1|, |R_2|\geq 2$. If $|R_2|=1$, we must have $|I_2|=1$ to avoid a relationship between momenta, and, similarly, if $|I_1|=1$, then $|R_1|=1$; so it only remains to consider a situation in which either $|I_2|=1$ or $|R_1|=1$, the cases being similar. 

Suppose $|R_1|=1$; if $|I_2|, |R_2|\geq 3$, the above arguments shows that $I_2, R_1$ and $R_2$ are consecutive. This leaves the cases where $|R_2|=2$, which is an NMHV amplitude, and $|I_2|=2$. In either case, it is straightforward to  show by explicit calculation that the only contribution comes when the sets are consecutive. 

If $c_{i_mr_1}=0$, so that $i_m\in I_2, r_1\in R_1$,we can use (\ref{shift}) to rewrite the factors of $\C_{ab}$ successively in (\ref{ResMC4}),
starting with $\C_{2,n-1}$ 
\be
\C_{2,n-1}\rightarrow\left[{c^{i_2i_{3}}_{r_{n-2}r_{n-1}}c_{i_{1}r_{n-2}}c_{i_{3}r_{n}}\over c^{i_2i_m}_{r_1r_{n-1}}c_{i_{1}r_1}c_{i_mr_{n}}}\right]\C^{i_mi_1i_2}_{r_1r_{n-1}r_{n}}
\rightarrow -c^{i_1i_{2}}_{r_{n-1}r_{n}}c^{i_2i_{3}}_{r_{n-2}r_{n-1}}c_{i_{1}r_{n-2}}c_{i_{3}r_{n}},
\label{shift7}
\ee
and working towards $\C_{m-1,2}$, 
\be 
\prod_{a=2}^{m'} \prod_{b=n'+1}^{n-1} {1\over \C_{ab}}\rightarrow\prod_{a=2}^{m'} \prod_{b=n'+1}^{n-1} 
\left[-{1\over c^{i_ai_{a+1}}_{r_{b-1}r_b}c_{i_{a-1}r_{b-1}}c_{i_{a+1}r_{b+1}}}\right]
{1\over c^{i_{a-1}i_{a}}_{r_{b}r_{b+1}}},
\label{shift8}
\ee
$c_{i_mr_1}\rightarrow0$, giving 
\be \hbox{Res}_{z_{A}}\M_{mn}=K_{mn}\oint_{\overline\Oc'}  {G_{mn}\prod_{a=2}^{m-1}\prod_{b=2}^{n-1}c_{i_ar_b}d c_{i_ar_b}\over c_{i_1r_1}c_{i_mr_n}\prod_{a=1}^{m-1}\prod_{b=1}^{n-1}c^{i_ai_{a+1}}_{r_br_{b+1}}(z)}dz,\label{ResMC6}\ee
where, labeling $I_1,I_2, R_1, R_2$ as in  (\ref{IRC}), appropriate to case C,  the contour $\overline\Oc'$ is the sum of contours each encircling the zeros of
$c_{i_ar_b}, m'< a\leq m,\; 1\leq b\leq n'$ and $c^{i_{a}i_{a+1}}_{r_{b}r_{b+1}}(z), 1\leq a < m',\; n'<b<n$,
at least one of the factors $c^{i_{a-1}i_{a}}_{r_{b}r_{b+1}},c^{i_ai_{a+1}}_{r_{b-1}r_b},c_{i_{a-1}r_{b-1}},c_{i_{a+1}r_{b+1}},$ for $ 2\leq a\leq m'-1,\;2\leq b\leq n',$ and $ m'< a\leq m-1,\;n'+2\leq b\leq n-1$. As in section \ref{TypeC}, we can show that (\ref{ResMC6}) vanishes unless $(m',n')$ lies on the `snake' line, defined in section \ref{FormF}. Thus the negative helicity contributions to $\M_{mn}$ can be written as a sum of contributions of the form (\ref{ResMC6}) with the residue taken at a combination of factors of the denominator. We hope to discuss this further elsewhere. Case D can be treated similarly.

\section{Conclusions}
\label{Conc}

In this paper, we have developed the techniques of \cite{DG3, DGsh} for calculating general gluon tree amplitudes, with $m$ positive helicities  and $n$ negative helicities, expressed in terms of link variables. Our analysis is based on the approach of writing the integrand function, $F_{mn}$, for a general amplitude in the terms of the ratio, $G_{mn}$, of that function to the corresponding function, $F^{\hbox{\tiny split}}_{mn}$, for the split helicity amplitude. The form of $G_{mn}$, which is independent of the choice of constraints, can be given in terms of simple rules described in terms of the particular sequence of positive and negative helicities, which can be conveniently encoded using a `snake' line on an $m\times n$ tabular diagram. The form of $F_{mn}$ is also quite simple if we use contiguous constraints, as in (\ref{defC}), generalizing \cite{DGsh}. 

Our main objective in this paper has been to establish directly the BCFW relation  \cite{BCFW} for all twistor string gluon tree amplitudes, using the link variable approach, so confirming the equivalence of these amplitudes to the corresponding amplitudes in $N=4$ super Yang Mills theory. Making a displacement in the momenta, linear in $z$, following \cite{BCFW}, which is equivalent to translating one of the link variables by $z$, to obtain the amplitude, $\M_{mn}(z)$, as a function of $z$, the BCFW relation is established by writing $\M_{mn}(z)$ as a sum over its poles. These poles arise from poles in $F_{mn}$ pinching the contour onto the singularities in the integrand coming from the zeros of the constraint functions, as well as some spurious contributions as in \cite{DG3}. 

The possible poles in $F_{mn}$ are simply listed in (\ref{rpoles}) and (\ref{spoles}). This leads to four classes of residues or contributions (\ref{caseA})--(\ref{caseD}) to the BCFW relation (\ref{BCFWE}), which provide all the necessary terms. The `snake' line plays a central role that precisely the right terms contribute as residues, and also that the functions $G_{mn}$ factorize at points corresponding to the poles into two functions associated with the subamplitudes. The demonstration, in section \ref{Abs}, that there are no other contributions completes a proof of the relation. 

The algebraic expressions for the integrands in this paper can be conveniently represented  and manipulated using diagrammatic techniques, extending those used in \cite{DGsh}, outlined in Appendix \ref{Diagrams}. These can also be used to provide expressions for gluon tree amplitudes as a sum of a suitable set of diagrams.

\section*{Acknowledgements}
We are grateful to Nima Arkani-Hamed, Jacob Bourjaily, Freddy Cachazo,
Jaroslav Trnka for discussions.
LD thanks the Institute for Advanced Study at Princeton for its hospitality.
LD was partially supported by the U.S. Department of Energy, Grant No. 
DE-FG02-06ER-4141801, Task A.
\vfill\eject
\appendix

\section{\bf Relations for $G_{mn}$}
\label{RelG}

(a) {\sl Expressions for $G_{mn}$.} The integrand for the amplitude (\ref{ointhF}) is determined by $G_{mn}$, the integrand divided by the integrand for the split-helicity case, $G_{mn}(c)=F_{mn}(c)/F^{\hbox{\tiny split}}_{mn}(c)$. To find an expression for $G_{mn}(c)$ it is convenient to re-express the link variables, $c_{ir}$,  in terms of the twistor string variables $k_\alpha, \rho_\alpha$.The constraints (\ref{defC}) are equivalent to the link variables being of the form \cite{DG3}
\be
c_{ir}={k_i\over k_r(\rho_i-\rho_r)}, \qquad i\in I,\quad r\in R.\label{ckrho}
\ee
In terms of these variables, 
\be
F_{mn}(c)= f(k,\rho)\prod_{\alpha=1}^N{1\over \rho_\alpha-\rho_{\alpha+1}}
\ee
where the function $f(k,\rho)$ is the same for each $N$-gluon amplitude, independently of the order of the helicities, and $\rho_{N+1}\equiv\rho_{1}$. Consider an amplitude, such as that specified following (\ref{Fsplit}), in which, cyclically, there are $p$ strings of both positive and negative helicities, with $\epsilon_\alpha=+1$ for $r_{n_{e-1}}+1\leq \alpha \leq i_{m_e} $ and $\epsilon_\alpha=-1$ for $i_{m_e}+1\leq \alpha \leq  r_{n_{e}}$, $1\leq e\leq p$. Here we identify the indices $\alpha$ and $e$ cyclically so that $\alpha\equiv \alpha+N$ and $e\equiv e+p$, and, similarly, $i_{a+m}=i_a$ and $r_{b+n}=r_b$. [In section \ref{FormF}, we assumed that the first helicity was positive, $i_1=1$, and the last negative, $r_n=N$, but here we shall allow for the other three possibilities as well.] For such an amplitude
\begin{align}
g(\rho)&\equiv\prod_{\alpha=1}^N{1\over \rho_\alpha-\rho_{\alpha+1}}\cr
&=
\prod_{e=1}^p{(\rho_{i_{m_e}}-\rho_{i_{m_e+1}})(\rho_{r_{n_e}}-\rho_{r_{n_e+1}})
\over (\rho_{i_{m_e}}-\rho_{r_{n_{e-1}+1}})(\rho_{r_{n_e}}-\rho_{i_{m_e+1}})}
\prod_{a=1}^m{1\over \rho_{i_a}-\rho_{i_{a+1}}}
\prod_{b=1}^n{1\over \rho_{r_b}-\rho_{r_{b+1}}}
\end{align}
and the corresponding function for the split-helicity case is
\be
g_{\hbox{\tiny sp}}(\rho)={(\rho_{i_{m}}-\rho_{i_{1}})(\rho_{r_{n}}-\rho_{r_{1}})
\over (\rho_{i_{m}}-\rho_{r_{1}})(\rho_{r_{n}}-\rho_{i_{1}})}
\prod_{a=1}^m{1\over \rho_{i_a}-\rho_{i_{a+1}}}
\prod_{b=1}^n{1\over \rho_{r_b}-\rho_{r_{b+1}}},
\ee
so that
\begin{align}
G_{mn}={g(\rho)\over g_{\hbox{\tiny sp}}(\rho)}
&={ (\rho_{i_{m}}-\rho_{r_{1}})(\rho_{r_{n}}-\rho_{i_{1}})\over (\rho_{i_{m}}-\rho_{i_{1}})(\rho_{r_{n}}-\rho_{r_{1}})}
\prod_{e=1}^p{(\rho_{i_{m_e}}-\rho_{i_{m_e+1}})(\rho_{r_{n_e}}-\rho_{r_{n_e+1}})
\over (\rho_{i_{m_e}}-\rho_{r_{n_{e-1}+1}})(\rho_{r_{n_e}}-\rho_{i_{m_e+1}})}\label{G1}\\
&={ c_{i_1r_1}c_{i_mr_n}\over c^{i_1i_m}_{r_1r_n}}
\prod_{e=1}^pc_{i_{m_e}r_{n_{e-1}+1}}\prod_{e=1}^p{c^{i_{m_e}i_{m_e+1}}_{r_{n_e}r_{n_e+1}}
\over c_{i_{m_e}r_{n_e}}c_{i_{m_e}r_{n_e+1}}c_{i_{m_e+1}r_{n_e+1}}}.\label{G2}
\end{align}
This expression for $G_{mn}$ holds independently of the signs of the first and last helicities.

If the first helicity is positive and the last negative, as in section \ref{FormF}, we may take $r_{n_p}= r_n=N$, $r_{n_p+1}= r_1$, $i_{m_p}=i_m$, and $i_{m_p+1}\equiv i_1=1$. Then, the $e=p$ term in the last product in (\ref{G2}) largely cancels the first term on  the right hand side, implying (\ref{defG}),
\be
G_{mn}={1\over c_{i_mr_1}}\prod_{e=1}^pc_{i_{m_e}r_{n_{e-1}+1}}
\prod_{e=1}^{p-1}{c^{i_{m_e}i_{m_e+1}}_{r_{n_e}r_{n_e+1}}
\over c_{i_{m_e}r_{n_e}}c_{i_{m_e}r_{n_e+1}}c_{i_{m_e+1}r_{n_e+1}}}.\label{Gpn}
\ee
If both the first and the last helicities are positive, we may arrange that the adjacent indices $i_m,i_1$ are contained in the sequence of adjacent indices $i_{m_p+1},\ldots,i_{m_1}$, then $r_{n_p}= r_n$, $r_{n_p+1}= r_1$, and
\be
G_{mn}={ c_{i_1r_1}c_{i_mr_n}c^{i_{m_p}i_{m_p+1}}_{r_{n}r_{1}}\over c^{i_1i_m}_{r_1r_n}c_{i_{m_p}r_{n}}c_{i_{m_p}r_{1}}c_{i_{m_p+1}r_{1}}}
\prod_{e=1}^{p}c_{i_{m_e}r_{n_{e-1}+1}}\prod_{e=1}^{p-1}{c^{i_{m_e}i_{m_e+1}}_{r_{n_e}r_{n_e+1}}
\over c_{i_{m_e}r_{n_e}}c_{i_{m_e}r_{n_e+1}}c_{i_{m_e+1}r_{n_e+1}}}.\label{Gpp}
\ee
If both the first and the last helicities are negative, we may arrange that the adjacent indices $r_n,r_1$ are contained in the sequence of adjacent indices $r_{n_{p-1}+1},\ldots,r_{n_p}$, then $i_{m_p}= i_m$, $i_{m_p+1}= i_1$, and
\be
G_{mn}={ c_{i_1r_1}c_{i_mr_n}c^{i_{m}i_{1}}_{r_{n_p}r_{n_p+1}}\over c^{i_1i_m}_{r_1r_n}c_{i_{m}r_{n_p}}c_{i_{m}r_{n_p+1}}c_{i_{1}r_{n_p+1}}}
\prod_{e=1}^{p}c_{i_{m_e}r_{n_{e-1}+1}}\prod_{e=1}^{p-1}{c^{i_{m_e}i_{m_e+1}}_{r_{n_e}r_{n_e+1}}
\over c_{i_{m_e}r_{n_e}}c_{i_{m_e}r_{n_e+1}}c_{i_{m_e+1}r_{n_e+1}}}.\label{Gnn}
\ee
If the first helicity is negative and the last positive, we may obtain a simpler expression for $G_{mn}$, comparable to (\ref{Gpn}), by first writing
\begin{align}
G_{mn}&={ (\rho_{i_{m}}-\rho_{r_{1}})(\rho_{r_{n}}-\rho_{i_{1}})\over (\rho_{i_{m}}-\rho_{i_{1}})(\rho_{r_{n}}-\rho_{r_{1}})}
\prod_{e=1}^p{(\rho_{i_{m_e}}-\rho_{i_{m_e+1}})(\rho_{r_{n_{e-1}}}-\rho_{r_{n_{e-1}+1}})
\over (\rho_{i_{m_e}}-\rho_{r_{n_{e-1}+1}})(\rho_{r_{n_e}}-\rho_{i_{m_e+1}})}\label{G3}\\
&={ c_{i_1r_1}c_{i_mr_n}\over c^{i_1i_m}_{r_1r_n}}
\prod_{e=1}^pc_{i_{m_e+1}r_{n_{e}}}\prod_{e=1}^p{c^{i_{m_e}i_{m_e+1}}_{r_{n_{e-1}}r_{n_{e-1}+1}}
\over c_{i_{m_e}r_{n_{e-1}}}c_{i_{m_e+1}r_{n_{e-1}}}c_{i_{m_e+1}r_{n_{e-1}+1}}},\label{G4}
\end{align}
which provides an alternative expression to (\ref{G2}) for $G_{mn}$ holding independently of the signs of the first and last helicities. If the first helicity is negative and the last positive, we may take $i_{m_p}= i_m=N$, $i_{m_p+1}= i_{1}$, $r_{n_{p-1}}=r_n$, and $r_{n_{p-1}+1}=r_1$. Then,
\be
G_{mn}={ 1\over c_{i_{1}r_{n}}}
\prod_{e=1}^pc_{i_{m_e+1}r_{n_{e}}}\prod_{e=1}^{p-1}{c^{i_{m_e}i_{m_e+1}}_{r_{n_{e-1}}r_{n_{e-1}+1}}
\over c_{i_{m_e}r_{n_{e-1}}}c_{i_{m_e+1}r_{n_{e-1}}}c_{i_{m_e+1}r_{n_{e-1}+1}}}.\label{Gnp}
\ee
If the first helicity is positive and the last negative, as in section \ref{FormF}, again taking  $r_{n_p}= r_n=N$, $r_{n_p+1}= r_1$, $i_{m_p}=i_m$, and $i_{m_p+1}\equiv i_1=1$, then we may rewrite (\ref{G4}) as
\be
G_{mn}=k\;{ c^{i_{m_1}i_{m_1+1}}_{r_{1}r_{2}}c^{i_{m-1}i_{m}}_{r_{n_{p-1}}r_{n_{p-1}+1}}\over c^{i_{m-1}i_{m}}_{r_{1}r_{2}}}
\prod_{e=1}^{p-1}c_{i_{m_e+1}r_{n_{e}}}\prod_{e=2}^{p-1}{c^{i_{m_e}i_{m_e+1}}_{r_{n_{e-1}}r_{n_{e-1}+1}}
\over c_{i_{m_e}r_{n_{e-1}}}c_{i_{m_e+1}r_{n_{e-1}}}c_{i_{m_e+1}r_{n_{e-1}+1}}},\label{Gpn2}
\ee
where
\be
k={ c_{i_{m-1}r_{1}}c_{i_{m-1}r_{2}}c_{i_{m}r_{2}}\over c_{i_{m}r_{n_{p-1}}}c_{i_{m-1}r_{n_{p-1}}}c_{i_{m-1}r_{n_{p-1}+1}}
c_{i_{m_1}r_{2}}c_{i_{m_1+1}r_{2}}c_{i_{m_1+1}r_{1}}}.
\ee
To deduce (\ref{Gpn2}), we have used identities of the form \cite{DG3}
\be
{c^{ij}_{rs}c_{it}c_{jt}c_{iu}c_{ju}\over c^{ij}_{tu}c_{ir}c_{jr}c_{is}c_{js}}={k_tk_u(\rho_t-\rho_u)\over k_rk_s(\rho_r-\rho_s)}={c^{kl}_{rs}c_{kt}c_{lt}c_{ku}c_{lu}\over c^{kl}_{tu}c_{kr}c_{lr}c_{ks}c_{ls}}.
\ee

(b) {\sl Relation between $G, G^L$ and $G^R$ for Negative Helicity Contributions.} We derive the relation between $G_{mn}$ and the functions, $G_{ m',n'+1}^L,G_{ m''+1,n''}^R$, associated with the subamplitudes considered in section \ref{TypeC}. The corresponding functions $g(\rho), g^L(\rho), g^R(\rho)$ satisfy
\be
{g^L(\rho)g^R(\rho)\over g(\rho)}= {(\rho_\alpha-\rho_\beta)(\rho_{r_n}-\rho_{i_1})\over(\rho_\alpha-\rho_{r_{n'+1}})(\rho_{r_{n'+1}}-\rho_{i_1})
(\rho_{r_n}-\rho_{i_{m'}})(\rho_{i_{m'}}-\rho_{\beta})},
\ee
where $\alpha=i_{m'}$ or $r_{n'}$ and $\beta=i_{m'+1}$ or $r_{n'+1}$. The functions associated for the corresponding split amplitudes satisfy, 
\be
{g^L_{\hbox{\tiny sp}}(\rho)g^R_{\hbox{\tiny sp}}(\rho)\over g_{\hbox{\tiny sp}}(\rho)}= {(\rho_{r_n}-\rho_{i_1})(\rho_{i_m}-\rho_{r_1})\over
(\rho_{i_{m'}}-\rho_{r_1})(\rho_{r_{n'+1}}-\rho_{i_1})
(\rho_{r_n}-\rho_{i_{m'}})(\rho_{i_m}-\rho_{r_{n'+1}})}.
\ee
Then
\begin{align}
{G_{\sss m',n'+1}^LG_{\sss m''+1,n''}^R\over  G_{\sss mn}}&=
{g^L(\rho)g^R(\rho)g_{\hbox{\tiny sp}}(\rho)\over g(\rho)g^L_{\hbox{\tiny sp}}(\rho)g^R_{\hbox{\tiny sp}}(\rho)}\cr
&= 
{(\rho_\alpha-\rho_\beta)(\rho_{i_{m'}}-\rho_{r_1})
(\rho_{i_m}-\rho_{r_{n'+1}})\over(\rho_\alpha-\rho_{r_{n'+1}})
(\rho_{i_{m'}}-\rho_{\beta})(\rho_{i_m}-\rho_{r_1})}\cr
&= 
{c_{i_{m'}r_{n'+1}}c_{i_{m}r_{1}}\over c_{i_{m'}r_{1}}c_{i_{m}r_{n'+1}}}\times
\left[{(\rho_\alpha-\rho_\beta)(\rho_{i_m'}-\rho_{r_{n'+1}})\over 
(\rho_\alpha-\rho_{r_{n'+1}})
(\rho_{i_{m'}}-\rho_{\beta})}\right].\label{Gratio}
\end{align}
The second factor on the right hand side of (\ref{Gratio}) is unity for the permitted values of $\alpha$ and $\beta$, except for $\alpha=r_{n'}, \beta=i_{m'+1},$ when it equals $c_{i_{m'}r_{n'}}c_{i_{m'+1}r_{n'+1}}/c_{r_{n'}r_{n'+1}}^{i_{m'}i_{m'+1}}$, which also equals unity when $c_{i_{m'+1}r_{n'}}=0$, establishing (\ref{sG}),
\be
\left[{c_{i_mr_1}c_{i_{m'}r_{n'+1}}\over c_{i_{m'}r_1}c_{i_{m}r_{n'+1}}}\right]G_{\sss mn}
=G_{\sss m',n'+1}^LG_{\sss m''+1,n''}^R,\label{nGrel}
\ee
provided that $c_{i_{m'+1}r_{n'}}=0$, which holds for contributions of type C.

(c) {\sl Relation between $G, G^L$ and $G^R$ for Positive Helicity Contributions.} We now derive the relation between $G_{mn}$ and the functions, $G_{m'+1,n'}^L,G_{ m'',n''+1}^R$, associated with the subamplitudes considered in section \ref{TypeD}. The corresponding functions $g(\rho), g^L(\rho), g^R(\rho)$ satisfy
\begin{align}
{g^L(\rho)g^R(\rho)\over g(\rho)}&=
{(\rho_{\alpha}-\rho_{\beta})(\rho_{r_n}-\rho_{i_1})\over
(\rho_\alpha - \rho_{i_{m'+1}})(\rho_{i_{m'+1}}-\rho_{i_1})(\rho_{r_n}-
\rho_{r_{n'}}) (\rho_{r_{n'}}-\rho_\beta)}
\end{align}
where $\alpha=i_{m'}$ or $r_{n'}$ and $\beta=i_{m'+1}$ or $r_{n'+1}$. The functions associated for the corresponding split amplitudes satisfy, 
\begin{align}
{g^L_{\hbox{\tiny sp}}(\rho)g^R_{\hbox{\tiny sp}}(\rho)\over 
g_{\hbox{\tiny sp}}(\rho)} 
&= {(\rho_{r_n}-\rho_{i_1})(\rho_{i_m}-\rho_{r_1})\over
(\rho_{i_{m'+1}}-\rho_{r_1})(\rho_{r_{n'}}-\rho_{i_1})
(\rho_{r_n}-\rho_{i_{m'+1}})(\rho_{i_m}-\rho_{r_{n'}})}.
\end{align}
\begin{align}
&\hskip-12pt{G_{\sss m'+1,n'}^LG_{\sss m'',n''+1}^R\over  G_{\sss mn}}=
{g^L(\rho)g^R(\rho) g_{\hbox{\tiny sp}}(\rho)\over g(\rho)
g^L_{\hbox{\tiny sp}}(\rho)g^R_{\hbox{\tiny sp}}(\rho)}\cr
&\quad=
{(\rho_\alpha-\rho_\beta)(\rho_{i_{m'+1}}-\rho_{r_1})(\rho_{r_{n'}}-\rho_{r_1})
(\rho_{r_n}-\rho_{i_{m'+1}})
(\rho_{i_m}-\rho_{r_{n'}})\over(\rho_\alpha-\rho_{i_{m'+1}})(\rho_{i_{m'+1}}-
\rho_{i_1})(\rho_{r_n}-\rho_{r_{n'}})
(\rho_{r_{n'}}-\rho_{\beta})(\rho_{i_m}-\rho_{r_1})}\cr
&\quad= { c_{i_1r_n} c_{i_{m'+1}r_{n'}}\over c_{i_1r_{n'}}c_{i_{m'+1}r_{n'}}}
\left [{(\rho_\alpha-\rho_\beta)(\rho_{i_{m'+1}}-\rho_{r_1})
(\rho_{i_m}-\rho_{r_{n'}}) (\rho_{i_1}-\rho_{r_n})(\rho_{i_{m'+1}}-\rho_{r_{n'}})
\over (\rho_\alpha-\rho_{i_{m'+1}})(\rho_{i_{m'+1}}-
\rho_{i_1})(\rho_{r_n}-\rho_{r_{n'}})
(\rho_{r_{n'}}-\rho_{\beta})(\rho_{i_m}-\rho_{r_1}) }\right].\qquad\label{Gratio2}
\end{align}
For the permitted values of $\alpha$, $\beta$,
other than $\alpha = i_{m'},\beta= r_{n'+1}$, the second factor in 
(\ref{Gratio2}) equals
\be
{c_{i_mr_1} c_{i_{m'+1} r_n} c_{i_{m'+1}r_{n'}} c_{i_1r_{n'}}
\over c_{i_{m'+1}r_1} c_{i_mr_{n'}} c^{i_{m'+1}i_1}_{r_nr_{n'}}},\label{ccc}
\ee 
and
\begin{align}
\left.{c_{i_mr_1} c_{i_{m'+1} r_n} c_{i_{m'+1}r_{n'}} c_{i_1r_{n'}}
\over c_{i_{m'+1}r_1} c_{i_mr_{n'}} c^{i_{m'+1}i_1}_{r_nr_{n'}}}\right|_{c_{i_1r_n}
=0}
= {c_{i_mr_1} c_{i_{m'+1} r_n'}\over c_{i_mr_1} c_{i_{m'+1} r_n}},
\end{align}
which is unity for $c^{i_mi_{m'+1}}_{r_1r_{n'}}=0$. For 
$\alpha = i_{m'},\beta= r_{n'+1},$ the second factor in (\ref{Gratio2})
reduces to (\ref{ccc}) multiplied by 
\begin{align}
{c_{i_{m'}r_{n'}} c_{i_{m'+1}r_{n'+1}}\over 
c^{i_{m'}i_{m'+1}}_{r_{n'}r_{n'+1}}},
\end{align}
which is unity for $c_{i_{m'},r_{n'+1}}=0.$
Thus we have 
\begin{align}
\left[{c_{i_1r_n}c_{i_{m'+1}r_{n'}}\over c_{i_1r_{n'}}c_{i_{m'+1}r_n}}\right]
G_{\sss mn}
=G_{\sss m'+1,n'}^{L}G_{\sss m'',n''+1}^{R},\label{pGrel}
\end{align}
provided that $c_{i_1r_n}=c_{i_{m'}r_{n'+1}}=c^{i_mi_{m'+1}}_{r_1r_{n'}} =0$, which hold for contributions of type D.

\section{\bf Jacobians}
\label{Jacobians}

(a) {\sl Independent Variables.}
The momentum constraints  (\ref{L1z}),
\begin{align}
\pi_j &=\sum_{r\in R} c_{jr}\pi_r,\qquad
\bpi_{s} =-\sum_{i\in I}\bpi_{i} c_{is},\nonumber
\end{align}
can be used to express all the link variables $c_{ir},\; \; i\in I,\;  r\in R,$ in terms of the independent variables, $c_{i_ar_b}, \; 
2\leq a\leq m-1, \; \; 2\leq b\leq n-1.$ [We suppress the $z$ dependence in what follows.] In particular, 
\be
c_{i_1 r_b}=- {[i_m, r_b ]\over [i_m, i_1]}-\sum_{a=2}^{m-1} {[i_m, i_a ]\over [i_m, i_1]}c_{i_a r_b};
\qquad 
c_{i_m r_b}=- {[i_1, r_b ]\over [i_1, i_m]}-\sum_{a=2}^{m-1} {[i_1, i_a ]\over [i_1, i_m]}c_{i_a r_b};
\ee
\be
c_{i_ar_1}={\langle i_a, r_n\rangle\over \langle r_1,r_n\rangle}-\sum_{b=2}^{n-1} 
c_{i_ar_b}{\langle r_b, r_n\rangle\over \langle r_1,r_n\rangle}; \qquad 
c_{i_ar_n}={\langle i_a, r_1\rangle\over \langle r_n,r_1\rangle}-\sum_{b=2}^{n-1} 
c_{i_ar_b}{\langle r_b, r_1\rangle\over \langle r_n,r_1\rangle};
\ee
and
\be
c_{i_1 r_1}=- {[i_m, r_1 ]\over [i_m, i_1]}+{\langle i_1, r_n\rangle\over \langle r_1,r_n\rangle}- {[i_m|P| r_n\rangle\over[i_m, i_1] \langle r_1,r_n\rangle}
+\sum_{a=2}^{m-1}\sum_{r=2}^{n-1}  {[i_m, i_a ]\over [i_m, i_1]}c_{i_ar_b}{\langle r_b, r_n\rangle\over \langle r_1,r_n\rangle},
\ee
where $[i|P| r\rangle=\sum_{a=1}^m[i,i_a]\langle i_a,r\rangle$, and with similar relations for 
$c_{i_1 r_n}, c_{i_m r_1},$ and $c_{i_m r_n}$.

\vskip6pt

(b) {\sl Jacobian for Cases (A) and (B).} We calculate the Jacobian, $J^{A}$,  of 
\be
c^{i_1i_2}_{r_1r_2},c^{i_1i_2}_{r_2r_3},\dots, c^{i_1i_2}_{r_{n-2}r_{n-1}}\quad\hbox{with respect to}\quad
c_{i_2r_2}, c_{i_2r_3}, \ldots, c_{i_2r_{n-1}},
\ee
when 
\be
c^{i_1i_2}_{r_{b}r_{b+1}}=0, \qquad1\leq b\leq n-2,
\ee 
in which case we can write
\be
c_{i_{a}r_{b}}=\lambda_a\mu_b,\label{clm}
\ee
for $a=1,2$ and $1\leq b\leq n-1$.
\be
J^{A}_{bd}={\partial c^{i_1i_2}_{r_{b}r_{b+1}}\over\partial  c_{i_{2}r_{d+1}}}=
{[i_m, i_0 ]\over [i_m, i_1]}\left(\mu_{b}\delta_{bd}-\mu_{b+1}\delta_{b,d+1}+\mu_2{\langle r_{d+1}, r_n\rangle\over \langle r_1,r_n\rangle}\delta_{b1}\right),\quad 1\leq b,d\leq n-2;
\ee
where, as in (\ref{IR}), $\bpi_{i_0}=\bpi_{i_1}\lambda_1+\bpi_{i_2}\lambda_2$. 

Then
$\det J^{A}=\det \tilde J^{A},$ where
\be
\tilde J^{A}_{bd}=\sum_{e=d}^{n-2}{\mu_{e+1}\over\mu_{b+1}}J^{A}_{be}=
{[i_m, i_0 ]\over [i_m, i_1]}\left(\mu_{b}\delta_{bd}+\sum_{e=d}^{n-2}\mu_{e+1}{\langle r_{e+1}, r_n\rangle\over \langle r_1,r_n\rangle}\delta_{b1}\right)\ee
and thus
\begin{align}
\det J^{A}=\det\tilde J^{A}&= {[i_m, i_0]^{n-2}\over [i_m, i_1]^{n-2}}{\langle r_0, r_n\rangle\over \langle r_1,r_n\rangle}\prod_{b=2}^{n-2}\mu_b\cr
&={1\over\lambda_2^{n-2}\mu_1c_{i_2r_{n-1}}}  {[i_m, i_0]^{n-2}\over [i_m, i_1]^{n-2}}{\langle r_0, r_n\rangle\over \langle r_1,r_n\rangle}\prod_{b=1}^{n-1}c_{i_2r_b},\label{detA}
\end{align}
where, as in (\ref{IR}), $\pi_{r_0}=\sum_{b=1}^{n}\mu_b\pi_{r_b}$.

Similarly, for the Jacobian, $J^B$, of 
\be
c^{i_2i_3}_{r_{n-1}r_{n}},c^{i_3i_4}_{r_{n-1}r_{n}},\dots, c^{i_{m-1}i_m}_{r_{n-1}r_{n}}\quad\hbox{with respect to}\quad
c_{i_{2}r_{n-1}}, c_{i_{3}r_{n-1}}, \ldots, c_{i_{m-1}r_{n-1}},
\ee
when 
\be
c^{i_{a+1}i_{a+2}}_{r_{n-1}r_{n}}=0, \qquad1\leq a\leq m-2,
\ee 
so that (\ref{clm}) holds for $2\leq a\leq m$ and $ b=n-1,n$
\begin{align}
\det J^{B}&={[i_1, i_0]\over [i_1, i_m]}{\langle r_0, r_1\rangle^{m-2}\over \langle r_n,r_1\rangle^{m-2}}\prod_{a=3}^{m-1}\lambda_a\cr
&={1\over\lambda_m\mu_{n-1}^{m-2}c_{i_2r_{n-1}} }{[i_1, i_0]\over [i_1, i_m]}{\langle r_0, r_1\rangle^{m-2}\over \langle r_n,r_1\rangle^{m-2}}\prod_{a=2}^{m}c_{i_ar_{n-1}},\label{detB}
\end{align}
where now
$\bpi_{i_0}=\sum_{a=1}^m\bpi_{i_a}\lambda_a$ and $\pi_{r_0}=\mu_{n-1}\pi_{r_{n-1}}+\mu_{n}\pi_{r_{n}}$.

\vskip6pt

(c) {\sl Jacobian for Case (C).} We calculate the Jacobian, $J^{C}$,  of 
\be 
c_{i_ar_b},\; m'<a\leq m,\;1\leq b\leq n',\quad c^{i_{a-1}i_a}_{r_br_{b+1}},\; 
2\leq a\leq m',\; n'<b<n,\;(a,b)\ne (2,n-1),
\ee
with respect to
\begin{align}
c_{i_ar_b},\quad &m'< a< m,\; 2\leq b\leq n',\quad\hbox{\sl and }\quad a=m',\; 2\leq b\leq n',\cr
&\quad\hbox{\sl and }\quad m'< a< m,\;b=n'+1,\quad\hbox{\sl and }\quad 2\leq a\leq m',\; n'<b<n.
\end{align}
at $ c^{i_{a-1}i_{a}}_{r_{b}r_{b+1}}=0$ for $2\leq a\leq m',\; n'<b<n,\;(a,b)\ne (2,n-1),$
and $c_{i_ar_b}=0$ for $m'<a\leq m,\;1\leq b\leq n'$. This implies
\be
c_{i_ar_b}=\lambda_a\mu_b\qquad\hbox{for}\quad  1\leq a\leq m',\; n'<b\leq n,\;(a,b)\ne (1,n).\label{clmC}
\ee

For $2\leq c<m$, $2\leq d<n$, 
\be 
{\partial c_{i_ar_b}\over \partial c_{i_cr_d}}=\delta_{ac}\delta_{bd},\quad m'< a< m,\; 2\leq b\leq n';
\ee
\be
{\partial c_{i_{m}r_b}\over \partial c_{i_cr_d}}=-{[i_1,i_c]\over [i_1,i_m]}\delta_{bd},\quad  2\leq b\leq n';\quad
{\partial c_{i_ar_1}\over \partial c_{i_cr_d}}=-{\langle r_d,r_n\rangle\over \langle r_1,r_n\rangle}
\delta_{ac},\quad m'< a< m.
\ee

From this it follows that
\be
\det J^C= {[i_1,i_{m'}]^{n'-1}\langle r_{n'+1},r_{n}\rangle^{m''-1}\over[i_1,i_m]^{n'-1}\langle r_{1},r_{n}\rangle^{m''-1}}\det\hat J^C
\ee
where the Jacobian matrix $\hat J^C$ is defined by 
\begin{align}
\hat J^C_{ab,cd}&={\partial c^{i_{a-1}i_{a}}_{r_{b}r_{b+1}}\over\partial  c_{i_{c}r_{d}}},\quad
2\leq a\leq m',\; n'<b<n,\;(a,b)\ne (2,n-1),\cr
&=\left[\lambda_{a}\delta_{a-1,c}-\lambda_{a-1}\delta_{ac}-{[i_m,i_c]\over [i_m,i_1]}\lambda_2\delta_{a2}\right]
\left[\mu_{b+1}\delta_{bd}-\mu_{b}\delta_{b+1,d}+{\langle r_d, r_1\rangle\over \langle r_n,r_1\rangle}\mu_{n-1}\delta_{b,n-1}\right],\cr
\hat J^C_{2\;n-1,cd}&={\partial c_{i_{m}r_{1}}\over\partial  c_{i_{c}r_{d}}}
={[i_1,i_c]\over [i_1,i_m]}{\langle r_d, r_n\rangle\over \langle r_1,r_n\rangle}
\end{align}
for $2\leq c\leq m',\; n'<d<n$.

\begin{align}
{\lambda_{c}\mu_d\over \lambda_{a}\mu_{b}}\hat J^C_{ab,cd}&=\left[\lambda_{a-1}(\delta_{a-1,c}-\delta_{ac})-{[i_m,i_c]\over [i_m,i_1]}\lambda_c\delta_{a2}\right]
\left[\mu_{b+1}(\delta_{bd}-\delta_{b+1,d})+{\langle r_d, r_1\rangle\over \langle r_n,r_1\rangle}\mu_{d}\delta_{b,n-1}\right]\nonumber
\end{align}
for $(a,b)\ne(2,n-1);$
\be
{\lambda_{c}\mu_d\over \lambda_{2}\mu_{n-1}}\hat J^C_{2\;n-1,cd}=
{\lambda_{c}\mu_d\over \lambda_{2}\mu_{n-1}}{[i_1,i_c]\over [i_1,i_m]}{\langle r_d, r_n\rangle\over \langle r_1,r_n\rangle}.
\ee
Then
\begin{align}
\tilde J^C_{ab,cd}&=\sum_{e=c}^{m'}\sum_{f=n'+1}^{d}{\lambda_{e}\mu_f\over \lambda_{a}\mu_{b}}\hat J^C_{ab,ef}
=\left[-\lambda_{a-1}\delta_{ac}-{[i_m,i_c']\over [i_m,i_1]}\delta_{a2}\right]
\left[\mu_{b+1}\delta_{bd}+{\langle r_d', r_1\rangle\over \langle r_n,r_1\rangle}\delta_{b,n-1}\right]
\nonumber
\end{align}
for $(a,b)\ne(2,n-1);$
\be
\tilde J^C_{2\;n-1,cd}=\sum_{e=c}^{m'}\sum_{f=n'+1}^{d}{\lambda_{e}\mu_f\over \lambda_{a}\mu_{b}}\hat J^C_{2\;n-1,cd}=
{1\over \lambda_{2}\mu_{n-1}}{[i_1,i_c']\over [i_1,i_m]}{\langle r_d', r_n\rangle\over \langle r_1,r_n\rangle},
\ee
where $i_c'$ and $r_d'$ stand for 
\be\bpi_{i_c}'=\sum_{e=m'}^{c} \lambda_e\bpi_{i_e},\qquad \pi_{r_d}'=\sum_{f=n'+1}^{d}\mu_f\pi_{r_f}.
\ee
The matrix $\tilde J^C_{ab,cd}$ is straightforward to diagonalize and so $\det \tilde J^C_{ab,cd}$ is seen to be the product of the diagonal elements:
\be 
\det \tilde J^C_{ab,cd}={[i_1,i_0]\langle r_0, r_n\rangle\over \lambda_{2}\mu_{n-1}}{[i_m,i_{0}]^{n''-2}\langle r_{0},r_{1}\rangle^{m'-2}\over[i_m,i_1]^{n''-1}\langle r_{n},r_{1}\rangle^{m'-1}}\prod_{a=2}^{m'-1}\lambda_{a}^{n''-1}\prod_{b=n'+2}^{n-1}\mu_{b}^{m'-1}
\ee
where, as in (\ref{IR}),
\be\bpi_{i_0}=\sum_{e=1}^{m'} \lambda_e\bpi_{i_e},\qquad \pi_{r_0}=\sum_{f=n'+1}^{n}\mu_f\pi_{r_f}.
\ee
Thus, using (\ref{iPr}), the Jacobian determinant is given by 
\begin{align}
\det J^C&=
{[i_1|P_{m'n'}|r_n\rangle[i_{0},i_m]^{n''-2}\langle r_{1},r_{0}\rangle^{m'-2}[i_1,i_{m'}]^{n'-1}\langle r_{n'+1},r_{n}\rangle^{m''-1}\over\lambda_{2}\mu_{n-1}[i_1,i_m]^{n-2}\langle r_{1},r_{n}\rangle^{m-2}}\prod_{a=2}^{m'-1}\lambda_{a}^{n''-1}\prod_{b=n'+2}^{n-1}\mu_{b}^{m'-1}\cr
&=
{[i_1|P_{m'n'}|r_n\rangle[i_{0},i_m]^{n''-2}\langle r_{1},r_{0}\rangle^{m'-2}[i_1,i_{m'}]^{n'-1}\langle r_{n'+1},r_{n}\rangle^{m''-1}\over c_{i_2r_{n-1}}\lambda_{m'}^{n''}\mu_{n'+1}^{m'}[i_1,i_m]^{n-2}\langle r_{1},r_{n}\rangle^{m-2}}c_{i_{m'}r_{n'+1}}\prod_{a=2}^{m'}\prod_{b=n'+1}^{n-1}c_{i_ar_b}\label{detC}
\end{align}
\vskip6pt

(d) {\sl Jacobian for Case (D).} We calculate the Jacobian, $J^{D}$,  of 
\be 
c_{i_ar_b},\; 1\leq a\leq m',\;n'< b\leq n,\;(a,b)\ne (1,n),\quad c^{i_{a}i_{a+1}}_{r_{b-1}r_{b}},\; 
m'< a< m,\; 2\leq b\leq n',\label{cD}
\ee
with respect to
\begin{align}
c_{i_ar_b},\quad &2\leq a\leq m',\;n'< b< n,\quad\hbox{\sl and }\quad a=m'+1,\; n'< b< n,\cr
&\quad\hbox{\sl and }\quad 2\leq a\leq m',\;b=n',\quad\hbox{\sl and }\quad m'< a< m,\; 2\leq b\leq n'.
\end{align}
at $ c^{i_{a}i_{a+1}}_{r_{b-1}r_{b}}=0$ for $m'< a< m,\; 2\leq b\leq n',$
and $c_{i_ar_b}=0$ for $1\leq a\leq m',\;n'< b\leq n,\;(a,b)\ne (1,n)$. 
We first assume that $m'\leq 2$ and $n''\leq 2$ and deal with the remaining cases $(m',n')=(1,1)$ or $(m-1,n-1)$ later. The conditions $ c^{i_{a}i_{a+1}}_{r_{b-1}r_{b}}=0$ imply
\be
c_{i_ar_b}=\lambda_a\mu_b\qquad\hbox{for}\quad  m'< a\leq m,\; 1\leq b\leq n'.\label{clmD}
\ee

For $2\leq c<m$, $2\leq d<n$, 
\be 
{\partial c_{i_ar_b}\over \partial c_{i_cr_d}}=\delta_{ac}\delta_{bd},\quad 2\leq a\leq m',\;n'< b< n;
\ee
\be
{\partial c_{i_{1}r_b}\over \partial c_{i_cr_d}}=-{[i_m,i_c]\over [i_m,i_1]}\delta_{bd},\quad  n'< b< n;\quad
{\partial c_{i_ar_n}\over \partial c_{i_cr_d}}=-{\langle r_d,r_1\rangle\over \langle r_n,r_1\rangle}
\delta_{ac},\quad 1< a\leq m'.
\ee

From this it follows that
\be
\det J^D= {[i_m,i_{m'+1}]^{n''-1}\langle r_{n'},r_{1}\rangle^{m'-1}\over[i_m,i_1]^{n''-1}\langle r_{n},r_{1}\rangle^{m'-1}}\det\hat J^D
\ee
where the Jacobian $\hat J^D$ is defined by 
\begin{align}
\hat J^D_{ab,cd}&={\partial c^{i_{a}i_{a+1}}_{r_{b-1}r_{b}}\over\partial  c_{i_{c}r_{d}}},\quad
m'< a< m,\; 2\leq b\leq n',\cr
&=\left[\lambda_{a}\delta_{a+1,c}-\lambda_{a+1}\delta_{ac}-{[i_1,i_c]\over [i_1,i_m]}\lambda_{m-1}\delta_{a,m-1}\right]
\left[\mu_{b-1}\delta_{bd}-\mu_{b}\delta_{b-1,d}+{\langle r_d, r_n\rangle\over \langle r_1,r_n\rangle}\mu_{2}\delta_{b2}\right]\cr
\end{align}
\vskip-26pt
for $m'< c< m,\; 2\leq d\leq n'$.

\begin{align}
{\lambda_{c}\mu_d\over \lambda_{a}\mu_{b}}\hat J^D_{ab,cd}&=\left[\lambda_{a+1}(\delta_{a+1,c}-\delta_{ac})-{[i_1,i_c]\over [i_1,i_m]}\lambda_c\delta_{a,m-1}\right]
\left[\mu_{b-1}(\delta_{bd}-\delta_{b-1,d})+{\langle r_d, r_n\rangle\over \langle r_1,r_n\rangle}\mu_{d}\delta_{b2}\right]
\end{align}

Then
\begin{align}
\tilde J^D_{ab,cd}&=\sum_{e=m'+1}^{c}\sum_{f=d}^{n'}{\lambda_{e}\mu_f\over \lambda_{a}\mu_{b}}\hat J^C_{ab,ef}
=\left[-\lambda_{a+1}\delta_{ac}-{[i_1,i_c']\over [i_1,i_m]}\delta_{a,m-1}\right]
\left[\mu_{b-1}\delta_{bd}+{\langle r'_d, r_n\rangle\over \langle r_1,r_n\rangle}\delta_{b2}\right]
\end{align}
where $i_c'$ and $r_d'$ stand for 
\be\bpi_{i_c}'=\sum_{e=m'+1}^{c} \lambda_e\bpi_{i_e},\qquad \pi_{r_d}'=\sum_{f=d}^{n'}\mu_f\pi_{r_f}.
\ee
The matrix $\tilde J^D_{ab,cd}$ is straightforward to diagonalize and so $\det \tilde J^D_{ab,cd}$ is seen to be the product of the diagonal elements:
\be 
\det \tilde J^D_{ab,cd}={[i_1,i_{0}]^{n'-1}\langle r_{0},r_{n}\rangle^{m''-1}\over[i_1,i_m]^{n'-1}\langle r_{1},r_{n}\rangle^{m''-1}}\prod_{a=m'+2}^{m-1}\lambda_{a}^{n'-1}\prod_{b=2}^{n'-1}\mu_{b}^{m''-1}
\ee
\Black
where, as in (\ref{IR}),
\be\bpi_{i_0}=\sum_{e=m'+1}^{m} \lambda_e\bpi_{i_e},\qquad \pi_{r_0}=\sum_{f=1}^{n'}\mu_f\pi_{r_f}.
\ee
Again, using (\ref{iPr}),
\begin{align}
\det J^D&=
[i_1|P_{m_1n_1}|r_n\rangle {[i_{m'+1},i_m]^{n''-1}\langle r_{1},r_{n'}\rangle^{m'-1}[i_1,i_{0}]^{n'-2}\langle r_{0},r_{n}\rangle^{m''-2}\over[i_1,i_m]^{n-2}\langle r_{1},r_{n}\rangle^{m-2}}\prod_{a=m'+2}^{m-1}\lambda_{a}^{n'-1}\prod_{b=2}^{n'-1}\mu_{b}^{m''-1}\cr
&=
[i_1|P_{m_1n_1}|r_n\rangle {[i_{m'+1},i_m]^{n''-1}\langle r_{1},r_{n'}\rangle^{m'-1}[i_1,i_{0}]^{n'-2}\langle r_{0},r_{n}\rangle^{m''-2}\over\lambda_{m'+1}^{n' } \mu_{n'}^{m'' }[i_1,i_m]^{n-2}\langle r_{1},r_{n}\rangle^{m-2}}c_{m'+1, n'} \prod_{a=m'+1}^{m-1}\prod_{b=2}^{n'} c_{ab}.\label{detD}\end{align}

In the remaining cases, $(m',n')=(1,1)$ or $(m-1,n-1)$, the terms involving $c^{i_{a}i_{a+1}}_{r_{b-1}r_{b}}$ are absent from 
(\ref{cD}) and it is straightforward to calculate $\det J^D$, leading again to (\ref{detD}).

\section{\bf Diagrammatic Methods}
\label{Diagrams}
\setlength{\unitlength}{1.5mm}

We can use diagrams to represent, manipulate and organize the integrands that arise in the calculation of gluon tree amplitudes in twistor string theory. We denote the various factors that occur in the integrands diagrammatically by using a single circle for $c_{ir}$ , a double circle for $c_{ir}^2$, a triple circle for $c_{ir}^3$, and so on, a rectangle for $c^{ij}_{rs}$, and a solid disk for the constraint function $\C^{ijk}_{rst}$, suitably positioned on a grid whose rows are labeled by the positive helicities and the columns by negative helicities, as illustrated in Figure 2: 
\vskip12truemm

\hbox to \hsize{\hfil
\hbox{\hskip-25truemm
\begin{picture}(64,1)
\grid{-2}{0}{1}{1}
\put(-10,1){$i$}
\put(-7,5){$r$}
\puto{-1}{1}
\put(-2,2){$\equiv c_{ir};$}
\linethickness{0.075mm}
\multiput(10,0)(4,0){2}{\line(0,1){4}}
\multiput(10,0)(0,4){2}{\line(1,0){4}}
\put(8,1){$i$}
\put(11,5){$r$}
\thicklines
\put(12,2){\circle{1.5}}
\put(12,2){\circle{0.8}}
\put(16,2){$\equiv c_{ir}^2;$}
\grid{7}{0}{1}{1}
\put(26,1){$i$}
\put(29,5){$r$}
\putcc{8}{1}
\put(34,2){$\equiv c_{ir}^3;$}

\put(44,-0.8){$j$}\put(44,3.2){$i$}
\put(47.5,7){$r$}\put(51.5,7){$s$}
\linethickness{0.075mm}
\multiput(46,-2)(4,0){3}{\line(0,1){8}}
\multiput(46,-2)(0,4){3}{\line(1,0){8}}
\linethickness{0.3mm}
\multiput(48,0)(4,0){2}{\line(0,1){4}}
\multiput(48,0)(0,4){2}{\line(1,0){4}}
\put(56,2){$\equiv c^{ij}_{rs};$}
\grid{17}{-1}{3}{3}
\put(66,-3){$k$}\put(66,1){$j$}\put(66,5){$i$}
\put(70,9){$r$}\put(74,9){$s$}\put(78,9){$t$}
\const{19}{1}
\put(82,2){$\equiv \C^{ijk}_{rst}.$}
\thicklines
\Black
\put(37,-8){Figure 2}
\end{picture}}\hfil}
\vskip14truemm
Each of these can appear in either black, red, gold, and we shall use black $(\bullet)$ to denote factors in the numerator of an integrand, 
red $(\Red \bullet \Black)$ for factors in the denominator, and gold $(\Gold \bullet \Black)$
for factors in the denominator at which residues are being taken.  With this notation, the 
integral for the split-helicity amplitude given by (\ref{ointhF}) with $F_{mn}=F^{\hbox{\tiny split}}_{mn}$, defined by (\ref{Fsplit}), for  $m=7, n=8$, is described by the diagram,
\vskip42truemm

\hbox to \hsize{\hfil
\hbox{\hskip-25truemm
\begin{picture}(1,1)
\grid{-1}{-1}{8}{7}
\put(-7,21){$i_1$}\put(-7,17){$i_2$}\put(-7,13){$i_3$}\put(-7,9){$i_4$}
\put(-7,5){$i_5$}\put(-7,1){$i_6$}\put(-7,-3){$i_7$}
\put(-3,25){$r_1$}\put(1,25){$r_2$}\put(5,25){$r_3$}\put(9,25){$r_4$}
\put(13,25){$r_5$}\put(17,25){$r_6$}\put(21,25){$r_7$}\put(25,25){$r_8$}
\Gold
\multiput(2,18)(4,0){6}{\circle*{2.7}}
\multiput(2,14)(4,0){6}{\circle*{2.7}}
\multiput(2,10)(4,0){6}{\circle*{2.7}}
\multiput(2,6)(4,0){6}{\circle*{2.7}}
\multiput(2,2)(4,0){6}{\circle*{2.7}}
\Red
\dgrid{6}{0}{1}{1}
\dgrid{0}{5}{1}{1}
\Black
\dgrid{1}{1}{5}{4}
\thicklines
\multiput(2,22)(4,0){5}{\circle{1.5}}
\multiput(2,18)(4,0){5}{\circle{1.5}}\multiput(2,18)(4,0){5}{\circle{0.8}}
\multiput(2,14)(4,0){6}{\circle{1.5}}\multiput(2,14)(4,0){6}{\circle{0.8}}
\multiput(6,14)(4,0){4}{\circle{2}}
\multiput(2,10)(4,0){6}{\circle{1.5}}\multiput(2,10)(4,0){6}{\circle{0.8}}
\multiput(6,10)(4,0){4}{\circle{2}}
\multiput(2,6)(4,0){6}{\circle{1.5}}\multiput(2,6)(4,0){6}{\circle{0.8}}
\multiput(6,6)(4,0){4}{\circle{2}}
\multiput(6,2)(4,0){5}{\circle{1.5}}\multiput(6,2)(4,0){5}{\circle{0.8}}
\multiput(6,-2)(4,0){5}{\circle{1.5}}
\multiput(-2,6)(0,4){4}{\circle{1.5}}
\multiput(26,2)(0,4){4}{\circle{1.5}}
\puto{1}{1}\puto{6}{5}
\Black
\put(6,-8){Figure 3}
\end{picture}}\hfil}
\vskip18truemm
The corresponding diagram for the 15-point amplitude with helicites (\ref{15pt}), corresponding to the `snake' line illustrated in Figure 1, is given by multiplying $G_{mn}$, described by Figure 4a,  by the diagram of Figure 3, to give Figure 4b,
\vskip42truemm
\phantom{x}
\vskip22truemm
\hbox to \hsize{\hfil
\hbox{\hskip-25truemm
\begin{picture}(1,1)
\grid{-1}{-1}{8}{7}
\RedViolet
\linethickness{0.5mm}
\put(-4,12){\line(1,0){16}}\put(-4,12){\line(0,1){12}}
\put(12,4){\line(1,0){12}}\put(12,4){\line(0,1){8}}
\put(24,-4){\line(0,1){8}}\put(24,-4){\line(1,0){4}}
\Red
\puto{0}{0}\puto{3}{4}\puto{4}{3}\puto{4}{4}
\puto{6}{2}\puto{7}{2}\puto{7}{1}
\Black
\puto{0}{4}\puto{4}{2}\puto{7}{0}
\dgrid{3}{3}{1}{1}\dgrid{6}{1}{1}{1}
\put(33,9.5){$\rightarrow$}
\put(6,-8){Figure 4a}
\end{picture}}\hskip90truemm
\hbox{\hskip-25truemm
\begin{picture}(1,1)
\grid{-1}{-1}{8}{7}
\Gold
\multiput(2,18)(4,0){6}{\circle*{2.7}}
\multiput(2,14)(4,0){6}{\circle*{2.7}}
\multiput(2,10)(4,0){6}{\circle*{2.7}}
\multiput(2,6)(4,0){6}{\circle*{2.7}}
\multiput(2,2)(4,0){6}{\circle*{2.7}}
\Red
\dgrid{6}{0}{1}{1}\puto{0}{0}
\dgrid{0}{5}{1}{1}
\Black
\dgrid{1}{1}{5}{4}\dgrid{6}{1}{1}{1}
\thicklines
\multiput(2,22)(4,0){5}{\circle{1.5}}
\multiput(2,18)(4,0){5}{\circle{1.5}}\multiput(2,18)(4,0){5}{\circle{0.8}}
\multiput(2,14)(4,0){6}{\circle{1.5}}\multiput(2,14)(4,0){6}{\circle{0.8}}
\put(-2,14){\circle{0.8}}\put(6,14){\circle{2}}\put(18,14){\circle{2}}
\multiput(2,10)(4,0){6}{\circle{1.5}}\multiput(2,10)(4,0){6}{\circle{0.8}}
\put(6,10){\circle{2}}\put(10,10){\circle{2}}\put(18,10){\circle{2}}
\multiput(2,6)(4,0){6}{\circle{1.5}}\multiput(2,6)(4,0){5}{\circle{0.8}}
\multiput(6,6)(4,0){4}{\circle{2}}\put(14,6){\circle{2.7}}
\multiput(6,2)(4,0){5}{\circle{1.5}}\multiput(6,2)(4,0){5}{\circle{0.8}}
\multiput(6,-2)(4,0){6}{\circle{1.5}}
\multiput(-2,6)(0,4){4}{\circle{1.5}}
\multiput(26,10)(0,4){2}{\circle{1.5}}
\puto{1}{1}\puto{6}{5}
\put(10.3,10.3){\line(1,0){3.4}}\put(10.3,10.3){\line(0,1){3.4}}
\put(13.7,13.7){\line(-1,0){3.4}}\put(13.7,13.7){\line(0,-1){3.4}}
\Black
\put(6,-8){Figure 4b}
\end{picture}}
\hfil}
\vskip18truemm
The rules for $G_{mn}$ given before Figure 1 in section \ref{FormF} are represented in our diagrammatic notation by 
\vskip8truemm
\hbox{
\hbox{\hskip25truemm
\begin{picture}(1,1)
\grid{-1}{-1}{2}{2}
\RedViolet
\linethickness{0.5mm}
\put(0,-4){\line(0,1){4}}
\put(-4,0){\line(1,0){4}}
\Red
\puto{0}{1}\puto{1}{0}\puto{1}{1}
\Black
\dgrid{0}{0}{1}{1}
\put(-9,1){$ i_{a}$}\put(-9,-3){$i_{a+1}$}
\put(-3,5.5){$r_{b}$}\put(0,5.5){$r_{b+1}$}
\put(6,0){$\leftrightarrow$}
\end{picture}}
\hskip13truemm$\displaystyle
{c^{i_{a}i_{a+1}}_{r_{b}r_{b+1}}
\over c_{i_{a} r_{b}}c_{i_{a} r_{b+1}}c_{i_{a+1} r_{b+1}}};$
\hskip-2truemm
\hbox{\hskip18truemm
\begin{picture}(1,1)
\grid{-1}{-1}{2}{2}
\RedViolet
\linethickness{0.5mm}
\put(0,0){\line(0,1){4}}
\put(0,0){\line(1,0){4}}
\Black
\puto{1}{1}
\put(-9,1){$i_{a}$}\put(-9,-3){$i_{a+1}$}
\put(-3,5.5){$r_{b}$}\put(0,5.5){$r_{b+1}$}
\put(6,0){$\leftrightarrow$}
\put(-6,-8){Figure 5}
\end{picture}}
\hskip13truemm$\displaystyle c_{i_{a} r_{b+1}};
$
\hskip-1.5truemm
\hbox{\hskip18truemm
\begin{picture}(1,1)
\grid{-1}{-1}{2}{2}
\Red
\puto{0}{0}
\Black
\put(-9.5,1){$i_{m-1}$}\put(-9.5,-3){$i_{m}$}
\put(-3,5.5){$r_{1}$}\put(0,5.5){$r_{2}$}
\put(6,0){$\leftrightarrow$}
\end{picture}}
\hskip13truemm$\displaystyle {1\over c_{i_{m} r_{1}}}.$}

\vskip12truemm

Calculating the negative helicity contribution corresponding to the point on the `snake' line indicated in Figure 6a, following section \ref{TypeC}, involves taking residues as illustrated in Figure 6b,

\vskip36truemm
\hbox to \hsize{\hfil
\hbox{\hskip-25truemm
\begin{picture}(1,1)
\grid{-1}{-1}{8}{7}
\RedViolet
\linethickness{0.5mm}
\put(-4,12){\line(1,0){16}}\put(-4,12){\line(0,1){12}}
\put(12,4){\line(1,0){12}}\put(12,4){\line(0,1){8}}
\put(24,-4){\line(0,1){8}}\put(24,-4){\line(1,0){4}}
\Salmon
\put(12,8){\circle*{1.5}}
\Black
\put(6,-8){Figure 6a}
\end{picture}}\hskip90truemm
\hbox{\hskip-25truemm
\begin{picture}(1,1)
\grid{-1}{-1}{8}{7}
\Gold
\dgrid{4}{3}{3}{3}
\multiput(2,18)(4,0){3}{\circle*{2.7}}
\multiput(2,14)(4,0){3}{\circle*{2.7}}
\multiput(18,6)(4,0){2}{\circle*{2.7}}
\multiput(18,2)(4,0){2}{\circle*{2.7}}
\multiput(-2,6)(4,0){4}{\circle{1.5}}
\multiput(-2,2)(4,0){4}{\circle{1.5}}
\multiput(-2,-2)(4,0){4}{\circle{1.5}}
\Red
\Black
\put(6,-8){Figure 6b}
\end{picture}}
\hfil}
\vskip12truemm
which produces 
\vskip36truemm

\hbox to \hsize{\hfil
\hbox{\hskip-25truemm
\begin{picture}(1,1)
\grid{-1}{-1}{8}{7}
\Gold
\thicklines
\multiput(2,18)(4,0){3}{\circle*{2.7}}
\multiput(2,14)(4,0){3}{\circle*{2.7}}
\multiput(18,6)(4,0){2}{\circle*{2.7}}
\multiput(18,2)(4,0){2}{\circle*{2.7}}
\Red
\puto{4}{0}\puto{7}{2}
\dgrid{6}{0}{1}{1}\dgrid{4}{2}{1}{1}
\dgrid{0}{5}{1}{1}\puto{0}{3}
\RoyalBlue
\case{0}{3}{5}{4}
\Green
\case{4}{0}{4}{4}
\thicklines
\Black
\dgrid{5}{1}{2}{1}
\puto{7}{0}\puto{6}{0}\putc{6}{1}\puto{5}{1}\putc{5}{2}\putc{4}{2}\puto{5}{3}
\dgrid{1}{4}{2}{1}
\puto{2}{3}\puto{3}{3}\puto{0}{4}\puto{1}{4}\putc{2}{4}\puto{3}{4}
\puto{0}{5}\putc{1}{5}\putc{2}{5}\puto{3}{5}\puto{1}{6}\puto{2}{6}
\put(33.5,9.5){$\times$}
\put(6,-8){Figure 7a}
\end{picture}}
\hskip90truemm
\hbox{\hskip-25truemm
\begin{picture}(1,1)
\grid{-1}{-1}{8}{7}
\Gold
\dgrid{4}{3}{3}{3}
\thicklines
\multiput(-2,6)(4,0){4}{\circle{1.5}}
\multiput(-2,2)(4,0){4}{\circle{1.5}}
\multiput(-2,-2)(4,0){4}{\circle{1.5}}
\Red
\Black
\multiput(14,10)(4,0){3}{\circle{1.5}}
\multiput(14,14)(4,0){3}{\circle{1.5}}
\multiput(14,18)(4,0){3}{\circle{1.5}}
\put(14,10){\circle{0.8}}
\put(6,-8){Figure 7b}
\end{picture}}
\hfil}
\vskip15truemm
where the two subamplitudes into which the contribution factors are described by the subdiagrams in blue and green boxes in Figure 7a; the two subamplitudes  correspond to the `snake' lines illustrated by Figure 8a and Figure 8b. 

\vskip20truemm
\hbox to \hsize{\hfil
\hbox{\hskip-5truemm
\begin{picture}(1,1)
\grid{-1}{-1}{5}{4}
\RoyalBlue
\case{0}{0}{5}{4}
\RedViolet
\linethickness{0.5mm}
\put(-4,0){\line(1,0){16}}\put(-4,0){\line(0,1){12}}
\put(12,-4){\line(1,0){4}}\put(12,-4){\line(0,1){4}}
\Black
\put(0,-8){Figure 8a}
\end{picture}}
\hskip70truemm
\hbox{\hskip-15truemm
\begin{picture}(1,1)
\grid{-1}{-1}{4}{4}
\Green
\case{0}{0}{4}{4}
\RedViolet
\linethickness{0.5mm}
\put(-4,4){\line(1,0){12}}\put(-4,4){\line(0,1){8}}
\put(8,-4){\line(0,1){8}}\put(8,-4){\line(1,0){4}}
\Black
\put(-1,-8){Figure 8b}
\end{picture}}
\hfil}
\vskip15truemm
This corresponds to  an equation of the form of (\ref{contC}), 
\be
\hbox{Res}_{z_{4,4}}\M_{7,8}=-{z_{4,4}\over s_{4,4}}\M_{4,5}\M_{4,4}.\label{contC2}
\ee

We concluded in section \ref{Abs}  that negative helicity contributions can be written as sums of residues of (\ref{ResMC6}) and this has the diagrammatic representation 
\vskip42truemm

\hbox to \hsize{\hfil
\hbox{\hskip-25truemm
\begin{picture}(1,1)
\grid{-1}{-1}{8}{7}
\RedViolet
\linethickness{0.5mm}
\put(-4,12){\line(1,0){16}}\put(-4,12){\line(0,1){12}}
\put(12,4){\line(1,0){12}}\put(12,4){\line(0,1){8}}
\put(24,-4){\line(0,1){8}}\put(24,-4){\line(1,0){4}}
\Red
\puto{0}{0}\puto{3}{4}\puto{4}{3}\puto{4}{4}
\puto{6}{2}\puto{7}{2}\puto{7}{1}
\Black
\puto{0}{4}\puto{4}{2}\puto{7}{0}
\dgrid{3}{3}{1}{1}\dgrid{6}{1}{1}{1}
\put(33,9.5){$\times$}
\put(6,-8){Figure 4a}
\end{picture}}\hskip90truemm
\hbox{\hskip-25truemm
\begin{picture}(1,1)
\grid{-1}{-1}{8}{7}
\Red
\dgrid{6}{0}{1}{1}
\dgrid{0}{5}{1}{1}
\dgrid{0}{0}{7}{6}
\puto{7}{0}\puto{0}{6}
\Black
\multiput(2,18)(4,0){6}{\circle{1.5}}
\multiput(2,14)(4,0){6}{\circle{1.5}}
\multiput(2,10)(4,0){6}{\circle{1.5}}
\multiput(2,6)(4,0){6}{\circle{1.5}}
\multiput(2,2)(4,0){6}{\circle{1.5}}
\put(6,-8){Figure 9}
\end{picture}}
\hfil}

\vfill\eject

\vfill\eject

\singlespacing


\providecommand{\bysame}{\leavevmode\hbox to3em{\hrulefill}\thinspace}
\providecommand{\MR}{\relax\ifhmode\unskip\space\fi MR }
\providecommand{\MRhref}[2]
{
}
\providecommand{\href}[2]{#2}

\end{document}